\pdfoutput=1

\documentclass[%
 reprint,
 amsmath,amssymb,
 aps,
pra,
longbibliography,
]{revtex4-1}

\usepackage[T1]{fontenc}
\usepackage[latin9]{inputenc}
\usepackage{textcomp}
\usepackage{mathtools}
\usepackage{amsmath}
\usepackage{amssymb}
\usepackage{graphicx}

\makeatletter

\pdfpageheight\paperheight
\pdfpagewidth\paperwidth

\makeatother

\usepackage{babel}

\begin{document}
\title{Practical classical error correction for parity-encoded spin systems}
\author{Yoshihiro Nambu}
\affiliation{NEC-AIST Quantum Technology Cooperative Research Laboratory~~\\
 National Institute of Advanced Industrial Science and Technology }
\begin{abstract}
Quantum annealing (QA) has emerged as a promising candidate for fast solvers for combinatorial optimization problems (COPs) and has attracted the interest of many researchers. Since COP is logically encoded in the Ising interaction among spins, its realization necessitates a spin system with all-to-all connectivity, presenting technical challenges in the physical implementation of large-scale QA devices. W. Lechner, P. Hauke, and P. Zoller proposed a parity-encoding (PE) architecture consisting of an expanded spin system with only local connectivity among them to circumvent this difficulty in developing near-future QA devices. They suggested that this architecture not only alleviates implementation challenges and enhances scalability but also possesses intrinsic fault tolerance. This paper proposes a practical decoding method tailored to correlated spin-flip errors in spin readout of PE architecture. Our work is based on the close connection between PE architecture and classical low-density parity-check (LDPC) codes. We show that the bit-flip (BF) decoding algorithm can correct independent and identically distributed errors in the readout of the SLHZ system with comparable performance to the belief propagation (BP) decoding algorithm. Then, we show evidence that the proposed BF decoding algorithm can efficiently correct correlated spin-flip errors by simulation. The result suggests that introducing post-readout BF decoding reduces the computational cost of QA using the PE architecture and improves the performance of global optimal solution search. Our results emphasize the importance of the proper selection of decoding algorithms to exploit the inherent fault tolerance potential of the PE architecture.
\end{abstract}
\maketitle

\section{Introduction}

Combinatorial optimization problems (COPs) present significant mathematical challenges in various industrial applications, including routing, scheduling, planning, decision-making, transportation, and telecommunications. COPs have garnered the attention of many researchers, leading to intensive efforts to develop a fast solver for these problems. Since many COPs are NP-hard, effective approximation methods such as heuristics \cite{pearlHeuristicsIntelligentSearch1984} and meta-heuristics \cite{peresCombinatorialOptimizationProblems2021}, have often been employed to tackle related challenges. Some of these methods rely on probabilistic simulations of dynamic processes in physical systems. They are inspired by computational models of natural phenomena, which can be executed on digital computers or specially designed physical hardware.  Recently, there has been increasing interest in natural computing, with new nature-inspired computing hardware being proposed and analyzed based on intriguing natural systems such as neural networks, molecules, DNA, and quantum computers \cite{jiaoNatureInspiredIntelligentComputing2024}.

There is a growing interest in quantum annealing (QA) devices as fast solver candidates for COPs \cite{hengHowSolveCombinatorial2022}. The COPs can be mapped to a search for the ground state of the Hamiltonian of the Ising spin network. The QA device is designed to quickly search for such ground states by exploiting quantum phenomena. Various architectures of QA devices have been developed to solve large-scale industrial and social optimization problems in a reasonable amount of time. D-Wave Systems was the first to create a commercial QA device consisting of superconducting flux qubits \cite{johnsonQuantumAnnealingManufactured2011a,kingScalingAdvantagePathintegral2021,raymondHybridQuantumAnnealing2023,kingQuantumCriticalDynamics2023,kingCoherentQuantumAnnealing2022}. Another QA device, called a coherent Ising machine, has been developed using optical systems. \cite{wangCoherentIsingMachine2013,marandiNetworkTimemultiplexedOptical2014,mcmahonFullyProgrammable100spin2016,inagakiCoherentIsingMachine2016}. Kerr parametric oscillators (KPOs) have been proposed as alternative candidates for components of a QA device. \cite{gotoBifurcationbasedAdiabaticQuantum2016,niggRobustQuantumOptimizer2017a,puriQuantumAnnealingAlltoall2017a,zhaoTwoPhotonDrivenKerr2018,gotoQuantumComputationBased2019,onoderaQuantumAnnealerFully2020,gotoQuantumAnnealingUsing2020,kewmingQuantumCorrelationsKerr2020,kanaoHighaccuracyIsingMachine2021,yamajiCorrelatedOscillationsKerr2023}. 

To apply QA devices to a universal problem of COP, we need to simulate the fully connected graph model in a scalable manner using Ising spin hardware. This requirement is technically challenging, especially when implementing large-scale QA devices using on-chip solid-state technology, because many long-range interactions between spins are necessary. To alleviate this problem, we usually embed a network of logical spins with many long-range interactions in an enlarged network of physical spins that exhibit fewer long-range interactions. For example, a minor embedding technique was proposed that replaces long-range interactions between logical spins with short-range interactions between clusters of physical spins \cite{choiMinorembeddingAdiabaticQuantum2008a,choiMinorembeddingAdiabaticQuantum2011a}. In this technique, physical spins within the same cluster are connected in a chain by strong ferromagnetic interactions to act as a logical spin, and a minor-embedded Hamiltonian replaces the logical problem Hamiltonian. Strong ferromagnetic interactions act as energy penalties to align the orientation of the physical spins in a chain. This technique enables solving problems requiring fully connected graph structures by allowing them to be embedded into the sparse graph structures available for current QA hardware. 

However, it isn't easy to apply this technique to large-scale QA hardware scalably because it requires non-planar wiring between spins in this technique. In response, W. Lechner, P. Hauke, and P. Zoller (LHZ) proposed a parity encoding (PE) architecture, an embedded method that can be implemented scalably using planar circuitry \cite{lechnerQuantumAnnealingArchitecture2015}. 
In this architecture, physical spins are connected lattice-like by strong adjacent four-body interactions to simulate a fully connected logical spin system. The strong four-body interaction acts as an energy penalty to map a physical spin state to a logical spin state uniquely. As a result, both the logical local fields and logical couplings are mapped to the local fields that act on the physical spins. This offers a particularly attractive feature of the PE architecture: it can separate the problem of controlling local fields and couplings. Since the PE architecture encodes logical information redundantly and nonlocally in the physical degrees of freedom, LHZ suggested that this architecture includes an intrinsic fault tolerance. Later, F. Pastawski and J. Preskill (PP) \cite{pastawskiErrorCorrectionEncoded2016}  pointed out that the PE architecture is interpreted as a classical low-density parity-check (LDPC) code \cite{gallagerLowDensityParityCheckCodes1962,gallagerLowDensityParityCheckCodes1963} and demonstrated that if the spin readout errors are independent and identically distributed (i.i.d.), they can be corrected through the belief propagation (BP) algorithm, recognized as the standard decoding algorithm  for the LDPC codes \cite{pearlReverendBayesInference1982}. T. Albash, W. Vincl, and D. Lidar discussed whether several decoding algorithms can improve the performance of PE architectures \cite{albashSimulatedquantumannealingComparisonAlltoall2016}. 
They reported that when a simple majority vote decoding (MVD) is used for post-readout decoding, the PE architecture does not exhibit fault tolerance against the correlated spin-flip errors that arise in their model of QA. In addition, they showed that a more sophisticated minimum weight decoding (MWD) algorithm, while requiring significant effort over MVD, can boost decoding success over MVD for sufficiently weak four-body penalty constraints. Despite observing that the BP algorithm boosted decoding success substantially for sufficiently weak four-body penalty constraints, they concluded that this boost is an artifact only seen in special problem instances that depend on the principle of the MWD. They pointed out that for PE architecture to achieve performance comparable to or better than the ME method, it is essential to develop a better decoding strategy tailored to correlated spin-flip errors. Furthermore, it would be a matter of course that developing practical decoding algorithms with lower decoding costs is also essential to exploit the potential of the PE architecture.

This paper proposes a practical decoding algorithm applicable to QA devices based on PE architecture. The proposed algorithm is an iterative hard-decision decoding algorithm based on majority voting in a generalized syndrome, known as Gallager's bit-flipping (BF) algorithm within the context of LDPC codes \cite{gallagerLowDensityParityCheckCodes1962,gallagerLowDensityParityCheckCodes1963}. Although the complexity class of the MWD is NP-hard, that of our BF algorithm is P, similar to the BP algorithm. Furthermore, the algorithm requires less computational effort and processing time than the BP algorithm, and can be implemented by a logic circuit. Our BF algorithm is devised based on the close connection between the PE architecture and the classical LDPC codes noted by PP. We show that the spin system initially proposed by Sourlas \cite{sourlasSoftAnnealingNew2005} and later by LHZ (referred to as the SLHZ system) is a realization of the PE architecture, and COP based on the Hamiltonian of the SLHZ system is itself equivalent to decoding the LDPC codes. Assuming the i.i.d. noise model, we demonstrate that our BF decoding algorithm can correct spin readout errors with comparable convergence to the BP algorithm. 

To verify the potential for correcting correlated spin-flip errors in the BF decoding algorithm, we performed classical simulations of stochastically sampled spin readouts in the SLHZ system using Markov chain Monte Carlo (MCMC) sampling. This analysis allows us to study the tolerance to leakage errors due to purely classical thermal noise. We show evidence that by making the four-body penalty constraint sufficiently weak, the BF decoding algorithm, similar to the BP decoding algorithm, can efficiently correct correlated spin-flip errors in simulated readouts of SLHZ systems. We further show that applying BF decoding to the MCMC sampler's readouts reduces the computational cost in the MCMC sampling. Our findings suggest that a hybrid computational approach combining two different types of decoding algorithms, decoding by MCMC sampling and subsequent BF decoding, can mitigate the  calculation overhead inherent in the SLHZ systems. Our results were similar to those of the pioneering work by Albash et al., who used simulated quantum annealing to simulate the readout of the spins in the SLHZ system and used BP decoding to decode it. We sought to understand this phenomenon by comparing our algorithm with various known BF decoding algorithms. According to our analysis, this is a consequence of a combination of the Hamiltonian of the SLHZ system and the property common to the post-readout BF or BP decodings. Our BF decoding is promising for realizing near-term QA devices based on PE architecture.

This paper is organized as follows. Section \ref{sec:2}  briefly explains error-correcting codes (ECC) and probabilistic decoding as a preliminary step. Section \ref{sec:3} describes the close connection between LDPC codes and SLHZ systems, and proposes a simple BF decoding algorithm for SLHZ systems. Section \ref{sec:4} demonstrates the performance of the proposed BF decoding. In Sec.\ref{sec:5}, we outline several types of BF decoding algorithms and their relation to our algorithm. We explain why two-stage hybrid decoding performs better than the two algorithms individually. We also discuss relevance of our study to the Albash et al.'s work \cite{albashSimulatedquantumannealingComparisonAlltoall2016}. Section \ref{sec:6} concludes this paper.

\section{Preliminaries\label{sec:2}}

\subsection{Model}

\begin{figure*}
\includegraphics[viewport=150bp 140bp 800bp 450bp,clip,scale=0.65]{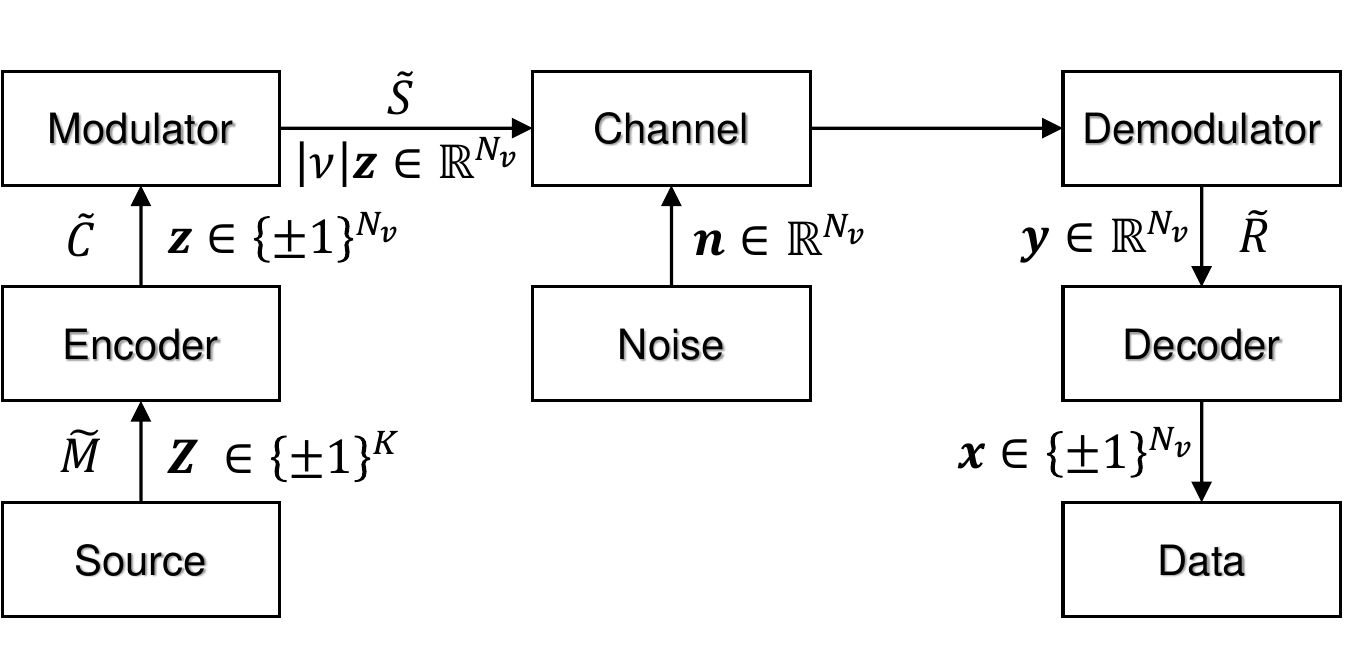}
\caption{A considered model for a communication system.\label{fig:1}}
\end{figure*}

First, let's describe our model. A binary source-word $\tilde{M}$ is first encoded into a binary code-word $\tilde{C}$ using ECC. The code-word is then modulated into the real signal $\tilde{S}$ and transmitted. During transmission, the signal is affected by noise in a transmission channel. At the end of the channel, the received signal is demodulated to obtain the observations $\tilde{R}$ (Fig.\ref{fig:1}). The ECC aims to communicate reliably over a noisy channel. Let $\tilde{M}$ and $\tilde{C}$ be $K$ bits and $N_{v}\left(>K\right)$ bits, respectively, and denote them by the binary vectors $\boldsymbol{\bar{Z}}=\left(\bar{Z}_{1},\ldots,\bar{Z}_{K}\right)\in\left\{ 0,1\right\} ^{K}$ and $\bar{\boldsymbol{z}}=\left(\bar{z}_{1},\ldots,\bar{z}_{N_{v}}\right)\in\left\{ 0,1\right\} ^{N_{v}}$, respectively. A Linear code is defined by a one-to-one map from $\tilde{M}$ in the set $\left\{ \bar{\boldsymbol{Z}}\right\} $ of $2^{K}$ source-words $\tilde{M}$ of length $K$ to $\tilde{C}$ in the set $\left\{ \bar{\boldsymbol{z}}\right\} $ of $2^{K}$ code-words $\tilde{C}$ of length $N_{v}$. They are specified by a generating matrix $\boldsymbol{G}_{K\times N_{v}}$ or a generalized parity check matrix $\boldsymbol{H}_{N_{c}\times N_{v}}$ satisfying $\boldsymbol{G}\boldsymbol{H}^{T}=\boldsymbol{0}_{K\times N_{c}}\:\left(\mathrm{mod}\:2\right)$. Here $\boldsymbol{G}$ and $\boldsymbol{H}$ are binary matrices (i.e., their elements are 0 or 1) and ``$\mathrm{mod}\:2$'' denotes that the multiplication is modulo two. The source-word $\boldsymbol{\bar{Z}}$ is mapped to the code-word $\bar{\boldsymbol{z}}$ by $\bar{\boldsymbol{z}}=\boldsymbol{\bar{Z}}\boldsymbol{G}\:\left(\mathrm{mod}\:2\right)$. Note that any $K$ linearly independent code-words can be used to form the generating matrix. For an arbitrary $N_{v}$-dimensional vector $\boldsymbol{\bar{x}}=\left(\bar{x}_{1},\ldots,\bar{x}_{N_{v}}\right)\in\left\{ 0,1\right\} ^{N_{v}}$, define the generalized syndrome vector $\boldsymbol{\bar{s}}\left(\boldsymbol{\bar{x}}\right)=\left(\bar{s}_{1}\left(\boldsymbol{\bar{x}}\right),\ldots,\bar{s}_{N_{c}}\left(\boldsymbol{\bar{x}}\right)\right)$ (generalized because it may not have $N_{v}-K$ bits) as $\boldsymbol{\bar{s}}\left(\boldsymbol{\bar{x}}\right)=\bar{\boldsymbol{x}}\boldsymbol{H}^{T}\:\left(\mathrm{mod}\:2\right)$. Then, $\bar{\boldsymbol{x}}$ is a code-word if and only if $\boldsymbol{\bar{s}}\left(\boldsymbol{\bar{x}}\right)=\boldsymbol{0}_{1\times N_{c}}$. This vector equation defines a set of generalized parity check equations consisting of a set of $N_{c}$ equations where only $N_{v}-K$ of them are linearly independent. It follows that the parity-check matrix for a given linear code can be chosen in many ways and that many syndrome vectors can be defined for the same code. The ratio of the length of the source-word to that of the code-word is called the rate: $r=K/N_{v}$. Later in this section, we will show specific examples of the matrices $\boldsymbol{G}$ and $\boldsymbol{H}$ for the PE architecture. The code-word \textbf{$\bar{\boldsymbol{z}}$} is converted into a sequence of bipolar variables $\boldsymbol{z}=\left(z_{1},\ldots,z_{N_{v}}\right)\in\left\{ \pm1\right\} ^{N_{v}}$ ($0$ is mapped to $+1$, and $1$ to $-1$) and modulated to antipodal signals $\left|v\right|\boldsymbol{z}$ by a binary phase shift keying modulation, where $\left|v\right|$ is the signal amplitude. Assume the signal $\tilde{S}$ is transmitted over a channel with additive white Gaussian noise (AWGN). At the end of the channel, the receiver obtains an observation $\tilde{R}$, denoted by an antipodal vector $\boldsymbol{y}=\left(y_{1},\ldots,y_{N_{v}}\right)=\left|v\right|\boldsymbol{z}+\boldsymbol{n}\in\mathbb{R}^{N_{v}}$ where $\boldsymbol{n}=\left(n_{1},\ldots,n_{N_{v}}\right)\in\mathbb{R}^{N_{v}}$ is a noise vector whose elements are i.i.d. Gaussian random variables with zero mean and variance $\sigma^{2}$. The goal of decoding is to reproduce the original source-word $\tilde{M}$ or associated code-word $\tilde{C}$ with a low bit error rate from the observation $\boldsymbol{y}$.

The link between the ECC and the spin glass model was first pointed out by Sourlas \cite{sourlasSpinglassModelsErrorcorrecting1989}. In the remainder of this paper, we will discuss our arguments primarily in the language of spin glass. Following Sourlas \cite{sourlasSpinGlassesErrorCorrecting1994,sourlasStatisticalMechanicsErrorcorrection1998,sourlasStatisticalMechanicsCapacityapproaching2001,sourlasnStatisticalMechanicsApproach2002}, our argument relies on isomorphism between the additive Boolean group $\left(\left\{ 0,1\right\} ,\oplus\right)$ and the multiplicative Ising group $\left(\left\{ \pm1\right\} ,\cdot\right)$, where a binary variable $\bar{a}{}_{i}\in\left\{ 0,1\right\} $ maps to the spin variable $a_{i}=\left(-1\right)^{\bar{a}_{i}}\in\left\{ \pm1\right\} $ and the binary sum maps to the product by $a_{i}a_{j}=(-1)^{\bar{a}_{i}\oplus\bar{a}_{j}}\in\left\{ \pm1\right\} $. Source-word $\bar{\boldsymbol{Z}}$ and code-word $\bar{\boldsymbol{z}}$ are mapped to vectors $\boldsymbol{Z}=\left(Z_{1},\ldots,Z_{K}\right)\in\left\{ \pm1\right\} ^{K}$ and $\boldsymbol{z}=\left(z_{1},\ldots,z_{N_{v}}\right)\in\left\{ \pm1\right\} ^{N_{v}}$ in the spin representation, respectively. Hereafter, we will refer to $\boldsymbol{Z}$ and $\boldsymbol{z}$ as a source-state and a code-state, respectively, and associated fictitious spins as logical and physical spins, respectively, following LHZ. Similarly, binary vectors $\bar{\boldsymbol{s}}$ and $\bar{\boldsymbol{x}}$ are mapped to the associated vectors $\boldsymbol{s}$ and $\boldsymbol{x}$ in the spin representation. In the following, we denote variables by symbols with and without an overbar in the binary and spin representations. It should be noted that by isomorphism, every addition of two binary variables corresponds to a unique product of spin variables and vice versa. For example, since $\bar{\boldsymbol{z}}=\boldsymbol{\bar{Z}}\boldsymbol{G}\:\left(\mathrm{mod}\:2\right)$ holds, $\boldsymbol{z}$ and $\boldsymbol{Z}$ are connected by the relation
\begin{equation}
z_{i}=\left(-1\right)^{\bigoplus_{j=1}^{K}\bar{Z}_{j}G_{ji}}=\prod_{\left\{ j:G_{ji}=1\right\} }Z_{j}\in\left\{ \pm1\right\} ,\label{eq:1}
\end{equation}
where $i=1,\ldots,N_{v}$. Similarly, $\boldsymbol{z}$ must satisfy the equation 
\begin{equation}
\left(-1\right)^{\bigoplus_{j=1}^{N_{v}}\bar{z}_{j}H_{ij}}=\prod_{\left\{ j:H_{ij}=1\right\} }z_{j}=\left(-1\right)^{0}=+1,\label{eq:2}
\end{equation}
for $i=1,\ldots,N_{c}$, since $\bar{\boldsymbol{z}}$ satisfies the parity check equation $\bar{\boldsymbol{z}}\boldsymbol{H}^{T}\:\left(\mathrm{mod}\:2\right)=\boldsymbol{0}_{1\times N_{c}}$. 

\subsection{Probabilistic decoding\label{subsec:2-B}}

Probabilistic decoding is performed based on statistical inference according to Bayes' theorem. To infer the code-state, consider the conditional probability $P\left(\boldsymbol{z}|\boldsymbol{y}\right)d\boldsymbol{y}$ that the prepared state is $\boldsymbol{z}$ when the observation was between $\boldsymbol{y}$ and $\boldsymbol{y}+d\boldsymbol{y}$. According to Bayesian inference, the Bayes optimal estimate is obtained when the posterior probability $P\left(\boldsymbol{z}|\boldsymbol{y}\right)$ or its marginals, discussed below, are maximized. According to Bayes' theorem, 
\begin{equation}
P\left(\boldsymbol{z}|\boldsymbol{y}\right)=\frac{P\left(\boldsymbol{y}|\boldsymbol{z}\right)P\left(\boldsymbol{z}\right)}{\sum_{\boldsymbol{z}}P\left(\boldsymbol{y}|\boldsymbol{z}\right)P\left(\boldsymbol{z}\right)}=\kappa P\left(\boldsymbol{y}|\boldsymbol{z}\right)P\left(\boldsymbol{z}\right)\label{eq:3}
\end{equation}
holds, where $\kappa$ is a constant independent of $\boldsymbol{z}$ which is determined by the normalization condition $\mathop{\sum_{\boldsymbol{z}}P\left(\boldsymbol{z}|\boldsymbol{y}\right)=1}$, and $P\left(\boldsymbol{z}\right)$ is the prior probability for the code-state $\boldsymbol{z}$. 

\subsubsection*{Word MAP decoding}

Sourlas showed that, based on Eqs.(\ref{eq:1}) and (\ref{eq:2}), two different formulations are possible for ECC. The first one is expressed in terms of source-state $\boldsymbol{Z}$. In this case, following Eq.(\ref{eq:1}), we assume the following prior probability for $\boldsymbol{z}$: 
\begin{equation}
P\left(\boldsymbol{z}\right)=\mu\prod_{i=1}^{N_{v}}\delta\left(z_{i},\prod_{\left\{ j:G_{ji}=1\right\} }Z_{j}\right),\label{eq:4}
\end{equation}
where $\mu$ is a normalization constant. The Kronecker's $\delta$ in Eq.(\ref{eq:4}) enforces the constraint that $\boldsymbol{z}$ obeys Eq.(\ref{eq:1}); it is a valid code-state. Assuming that the noise is independent for each spin and that $P\left(\boldsymbol{y}|\boldsymbol{z}\right)=\mathop{\underset{i=1}{\stackrel{N_{v}}{\prod}}P\left(y_{i}|z_{i}\right)}$ holds (memoryless channel), we can derive the following equation:
\begin{eqnarray}
-\ln P\left(\boldsymbol{z}|\boldsymbol{y}\right)
&=&\mathrm{const.}-\sum_{i=1}^{N_{v}}\theta\left(y_{i}\right)\prod_{\left\{ j:G_{ji}=1\right\} }Z_{j}\nonumber\\
&\equiv& H^{source}\left(\boldsymbol{Z}\right),
\label{eq:5}
\end{eqnarray}
where $\theta\left(y_{i}\right)$ is the half log-likelihood ratio (LLR) of the channel observation $y_{i}$, namely 
\begin{equation}
\theta\left(y_{i}\right)=\frac{1}{2}\log\frac{P\left(y_{i}|+1\right)}{P\left(y_{i}|-1\right)}.\label{eq:6}
\end{equation}

Alternatively, the second one is expressed in terms of code-state $\boldsymbol{z}$. In this case, following Eq.(\ref{eq:2}), we assume the following prior probability for $\boldsymbol{z}$: 
\begin{equation}
P\left(\boldsymbol{z}\right)=\mu\prod_{i=1}^{N_{c}} \delta\left(s_{i}\left(\boldsymbol{z}\right),+1\right),\label{eq:7}
\end{equation}
where
\begin{equation}
s_{i}\left(\boldsymbol{z}\right)=\prod_{\left\{ j:H_{ij}=1\right\} }z_{j}\label{eq:8}
\end{equation}
is the $i$th syndrome for $\boldsymbol{z}$  in the spin representation. Then, we can derive the following equation: 
\begin{eqnarray}
-\ln P\left(\boldsymbol{z}|\boldsymbol{y}\right)
&=&-\sum_{i=1}^{N_{v}}\theta\left(y_{i}\right)z_{i}+\underset{\gamma\rightarrow\infty}{\lim}\gamma\sum_{i=1}^{N_{c}}\frac{1-s_{i}\left(\boldsymbol{z}\right)}{2}\nonumber\\
&\equiv& H^{code}\left(\boldsymbol{z}\right).
\label{eq:9}
\end{eqnarray}
In Eq.(\ref{eq:9}), $\delta$'s in Eq.(\ref{eq:7}) is replaced by a soft constraint using the identity 
\begin{equation}
\delta\left(x,+1\right)=\underset{\gamma\rightarrow\infty}{\lim}\exp\left[-\gamma\frac{1-x}{2}\right].
\end{equation}
The second term of Eq.(\ref{eq:9}) enforces the constraint that $\boldsymbol{z}$ obeys Eq.(\ref{eq:2}); it is a valid code-state. In contrast, the first term reflects the correlation between the observation $\boldsymbol{y}$ and the code-state $\boldsymbol{z}$. 

Since a source-state corresponds one-to-one to an associated code-state, the two formulations above are equivalent as long as $\boldsymbol{z}$ is a code-state associated with a source-state $\boldsymbol{Z}$. It is evident that $H^{source}\left(\boldsymbol{Z}\right)$ is in the form of a spin glass Hamiltonian. In contrast, $H^{code}\left(\boldsymbol{z}\right)$ can be considered a Hamiltonian of an enlarged spin system. The state corresponding to the most probable word (``word maximum a posteriori probability'' or ``word MAP decoding''), i.e., the state that maximizes probability $P\left(\boldsymbol{z}|\boldsymbol{y}\right)$, is given by the ground state of the Hamiltonian $H^{source}\left(\boldsymbol{Z}\right)$ or $H^{code}\left(\boldsymbol{z}\right)$. In this case, probabilistic decoding corresponds to finding the most probable code-state $\boldsymbol{z}$, namely,
\begin{equation}
\boldsymbol{z}=\underset{\boldsymbol{x}\in C}{\arg\max}P\left(\boldsymbol{x}|\boldsymbol{y}\right)=\underset{\boldsymbol{x}\in C}{\arg\min}H^{code}\left(\boldsymbol{x}\right),
\label{eq:11}
\end{equation}where $C$ denotes the set of all code-states. Such decoded results can be obtained by, for example, simulated annealing or QA.

The LLR vector $\boldsymbol{\theta}\left(\boldsymbol{y}\right)=\left(\theta\left(y_{1}\right),\ldots,\theta\left(y_{N_{v}}\right)\right)\in\mathbb{R}^{N_{v}}$ is important because it contains all the information about the observation $\boldsymbol{y}$. Let be the likelihood for the additive white Gaussian noise (AWGN) channel as 
\begin{equation}
P\left(y_{i}|z_{i}\right)=\frac{1}{\sqrt{2\pi}\sigma}\exp\left[-\frac{\left(y_{i}-\left|v\right|z_{i}\right)^{2}}{2\sigma^{2}}\right],\label{eq:12}
\end{equation}
where $\left|v\right|$ and $\sigma^{2}$ are the amplitude of the prepared signal and the variance of the common Gaussian noise \cite{masseyThresholdDecoding1962}. Then the LLR is given by $\theta\left(y_{i}\right)=\beta y_{i}$, where $\beta=\tfrac{2\left|v\right|}{\sigma^{2}}>0$ is called the channel reliability factor, and its inverse  $\beta^{-1}$ is considered the temperature of the spin system in the language of spin glasses. Note that $\frac{1}{2}$$\left(\frac{\left|v\right|}{\sigma}\right)^{2}$ is the signal-to-noise ratio (SNR) of the AWGN channel. Thus, $\theta\left(y_{i}\right)$ is proportional to the magnitude of the channel observation $y_{i}$ for the AWGN channel. Here is a more detailed look at what LLR means \cite{masseyThresholdDecoding1962}. Suppose that the channel observation $y_{i}$ is hard-decided by $x_{i}=\mathrm{sign}\left[\theta\left(y_{i}\right)\right]$, where $\mathrm{sign}\left[\ldots\right]$ is the sign function. Then the error probability of this decision is given by 
\begin{equation}
\gamma_{i}=\frac{1}{1+e^{\left|\theta\left(y_{i}\right)\right|}}=\frac{e^{-\left|\theta\left(y_{i}\right)\right|}}{1+e^{-\left|\theta\left(y_{i}\right)\right|}}.\label{eq:13}
\end{equation}
Thus, the absolute value $\left|\theta\left(y_{i}\right)\right|$, called soft information, represents a reliability metric of the
hard decision $x_{i}=\mathrm{sign}\left[\theta\left(y_{i}\right)\right]$. It indicates how likely $x_{i}$ is to be $+1$ or $-1$. A value close to zero indicates low reliability, while a large value indicates high reliability. 

\subsubsection*{Symbol MAP decoding}

On the other hand, there is an alternative way for probabilistic decoding. Instead of considering the most probable word, it is also allowed to be interested only in the most probable symbol, i.e., the most probable value $z_{i}$ of the $i$th spin, ignoring the values of the other spin variables (``symbol MAP decoding''). In this strategy, we consider marginals 
\begin{eqnarray}
P\left(x_{i}|\boldsymbol{y}\right)
&:=&\sum_{x_{1}=-1}^{+1}\cdots\sum_{x_{i-1}=-1}^{+1}\sum_{x_{i+1}=-1}^{+1}\cdots\sum_{x_{N_{v}}=-1}^{+1}P\left(\boldsymbol{x}|\boldsymbol{y}\right)\nonumber\\
&=&\sum_{x_{k}\left(k\neq i\right)}P\left(\boldsymbol{x}|\boldsymbol{y}\right).
\end{eqnarray}
Decoding corresponds to finding the value of the spin variable that maximizes the marginals $P\left(x_{i}|\boldsymbol{y}\right)$, i.e.,
\begin{eqnarray}
\boldsymbol{z}_{i}
&=&\underset{x_{i}\in\left\{ \pm1\right\} }{\arg\max}P\left(x_{i}|\boldsymbol{y}\right)=\mathrm{sign}\left[\sum_{x_{i}=-1}^{+1}x_{i}P\left(x_{i}|\boldsymbol{y}\right)\right]\nonumber\\
&=&\mathrm{sign}\left[\left\langle x_{i}\right\rangle _{P\left(\boldsymbol{x}|\boldsymbol{y}\right)}\right].
\end{eqnarray}
A variety of algorithms can be used to achieve this task. For example, the best-known algorithm is the belief propagation (BP) algorithm for the LDPC codes \cite{pearlReverendBayesInference1982}. The BP
algorithm is an iterative algorithm where the LLR, initially given by the observation $\boldsymbol{y}$, is gradually increased in absolute values by taking account of parity check constraints. LLR is used in many decoding algorithms to measure reliability and uncertainty in binary random variables. The log of the likelihood is often used rather than the likelihood itself because it is easier to handle. Since the probability is always between $0$ and $1$, the log-likelihood is always negative, where larger values indicate a better-fitting model. 

In this work, an alternative simple algorithm is studied: the BF algorithm \cite{gallagerLowDensityParityCheckCodes1962,gallagerLowDensityParityCheckCodes1963}. This algorithm can be considered an approximation of the BP algorithm. It starts with an initial LLR $\boldsymbol{\theta}^{(0)}=\boldsymbol{\theta}\left(\boldsymbol{y}\right)$, and iteratively updates the hard decision $x_{i}^{(m)}=\mathrm{sign}\left[\theta_{i}^{(m)}\right]$ by a spin-flip operation determined by the parity check constraints and LLR $\boldsymbol{\theta}^{(m)}=\left(\theta_{1}^{(m)},\ldots,\theta_{n}^{(m)}\right)$ , which depends on the hard decision $\boldsymbol{x}^{(m-1)}$ of the previous round. A decoding algorithm that uses only information about the hard decision is called hard-decision decoding, while one that uses information about both the hard decision and its reliability metric is called soft-decision decoding. In the BF decoding, the reliability of the hard decision gradually increases as the bit-flipping is repeated. It is widely recognized that one of the key advantages of BF decoding over BP decoding is its simplicity and lower computational complexity. BF decoding is particularly useful in environments with limited computational resources because it requires fewer operations to decode the received message. This fact makes it a more efficient option when processing power is a concern. In contrast, although providing better error-correcting performance regarding bit error rate (BER), BP decoding typically involves more complex calculations and higher computational requirements. Therefore, BF decoding is often favored in applications where simplicity and low complexity are prioritized.

\section{Parity-encoding architecture and classical LDPC codes\label{sec:3}}

\subsection{Connection between SLHZ system and LDPC codes}

\begin{figure*}[tb]
\includegraphics[viewport=60bp 190bp 870bp 440bp,clip,scale=0.6]{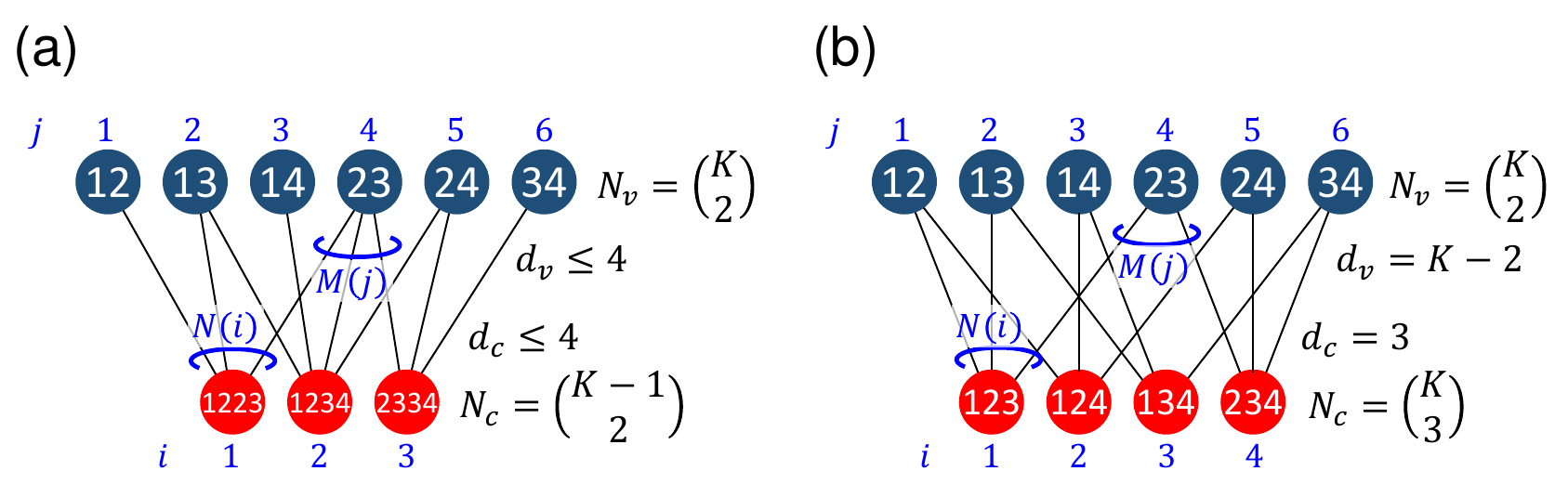}\caption{Two representations of bipartite graph for $K=4$ logical spins. The dark blue circle labeled $\left\{ k,l\right\} $ represents the variable $x_{kl}$, while the red circles labeled $\left\{ k,l,m,n\right\} $ or $\left\{ k,l,m\right\} $ represent the weight-4 syndrome $s_{klmn}^{(4)}$ and weight-3 syndrome $s_{klm}^{(3)}$, respectively. In these diagrams, let us relabel the variables with blue letters. An element of code-word vector $\boldsymbol{x}=\left(x_{1},\ldots,x_{N_{v}}\right)\in\left\{ +1,-1\right\} ^{N_{v}}$ is called variable node (VN). The $i$-th syndrome of $\boldsymbol{x}$ is defined by $s_{i}\left(\boldsymbol{x}\right)=\prod_{k\in N\left(i\right)}x_{k}\in\left\{ +1,-1\right\}$ and an element of vector  $\boldsymbol{s}(\boldsymbol{x})=\left(s_{1}\left(\boldsymbol{x}\right),\ldots,s_{N_{c}}\left(\boldsymbol{x}\right)\right)\in\left\{ +1,-1\right\} ^{N_{c}}$ is called a check node (CN), where $N\left(i\right)=\left\{ j:H_{ij}(H_{ij}^{'})=1\right\}$ is the VNs adjacent to a CN $i$ $\left(1\leq i\leq N_{c}\right)$ and $M\left(j\right)=\left\{ i:H_{ij}(H_{ij}^{'})=1\right\} $ is the CNs adjacent to a VN $j$ $\left(1\leq j\leq N_{v}\right)$. The column and row weights of the parity-check matrix are defined by $d_{v}(i)=\left|M\left(j\right)\right|$ and $d_{c}(i)=\left|N\left(i\right)\right|$, respectively. 
\label{fig:2}}
\end{figure*}

Next, consider an example of a PE architecture. The PE architecture corresponds to the following map: $\bar{z}_{ij}=\bar{Z}_{i}\oplus\bar{Z}_{j}$ for $1\leq i\leq j\leq k$. As a simple example, consider the case $K=4$ and assume that $\boldsymbol{\bar{Z}}=\left(\bar{Z}_{1},\bar{Z}_{2},\bar{Z}_{3},\bar{Z}_{4}\right)\in\left\{ 0,1\right\} ^{K}$ and $\bar{\boldsymbol{z}}=\left(\bar{z}_{12},\bar{z}_{13},\bar{z}_{14},\bar{z}_{23},\bar{z}_{24},\bar{z}_{34}\right)\in\left\{ 0,1\right\} ^{\tbinom{K}{2}}$. Suppose that the generating matrix is given by $K\times\tbinom{K}{2}$ matrix: 
\begin{equation}
\boldsymbol{G}=\begin{pmatrix}1 & 1 & 1 & 0 & 0 & 0\\
1 & 0 & 0 & 1 & 1 & 0\\
0 & 1 & 0 & 1 & 0 & 1\\
0 & 0 & 1 & 0 & 1 & 1
\end{pmatrix}.
\end{equation}
Note that there are two 1s in each column of $\boldsymbol{G}$. This is because each element of $\bar{\boldsymbol{z}}$ is the binary sum
of two elements of $\boldsymbol{\bar{Z}}$. We can consider the following two parity check matrices, 
\begin{equation}
\boldsymbol{H}=\begin{pmatrix}1 & 1 & 0 & 1 & 0 & 0\\
0 & 1 & 1 & 1 & 1 & 0\\
0 & 0 & 0 & 1 & 1 & 1
\end{pmatrix}
\end{equation} 
and 
\begin{equation}
\boldsymbol{H}'=\begin{pmatrix}1 & 1 & 0 & 1 & 0 & 0\\
1 & 0 & 1 & 0 & 1 & 0\\
0 & 1 & 1 & 0 & 0 & 1\\
0 & 0 & 0 & 1 & 1 & 1
\end{pmatrix},
\end{equation}
which are sparse matrices with mostly $0$s and relatively few $1$s when $K$ is very large. We can easily confirm that these parity-check matrices satisfy the constraints $\boldsymbol{G}\boldsymbol{H}^{T}=\boldsymbol{G}\boldsymbol{H}'{}^{T}=\boldsymbol{0}$. From these matrices, two different syndrome vectors can be derived. One is weight-4 syndrome vector $\boldsymbol{\bar{s}}^{(4)}(\bar{\boldsymbol{x}})=\left(\bar{s}_{1223}^{(4)},\bar{s}_{1234}^{(4)},\bar{s}_{2334}^{(4)}\right)=\bar{\boldsymbol{x}}\boldsymbol{H}^{T}\in\left\{ 0,1\right\} ^{\tbinom{K-1}{2}}$, and the other is weight-3 syndrome vector $\boldsymbol{\bar{s}}^{(3)}(\bar{\boldsymbol{x}})=\left(\bar{s}_{123}^{(3)},\bar{s}_{124}^{(3)},\bar{s}_{134}^{(3)},\bar{s}_{234}^{(3)}\right)=\bar{\boldsymbol{x}}\boldsymbol{H}'{}^{T}\in\left\{ 0,1\right\} ^{\tbinom{K}{3}}$, where $\bar{\boldsymbol{x}}=\left(\bar{x}_{12},\bar{x}_{13},\bar{x}_{14},\bar{x}_{23},\bar{x}_{24},\bar{x}_{34}\right)\in\left\{ 0,1\right\} ^{\tbinom{K}{2}}$
is an arbitrary binary vector. Only $2^{K}$ elements out of the $2^{\tbinom{K}{2}}$ possible $\bar{\boldsymbol{x}}$ are valid code-words satisfying the
parity check constraints $\boldsymbol{\bar{s}}^{(4)}(\bar{\boldsymbol{x}})=\boldsymbol{0}_{1\times \tbinom{K-1}{2}}$ or $\boldsymbol{\bar{s}}^{(3)}(\bar{\boldsymbol{x}})=\boldsymbol{0}_{1\times \tbinom{K}{3}}$. The connections between the variables $\bar{x}_{kl}$ and the weight-4 syndromes $\bar{s}_{klmn}^{(4)}$ are depicted by a sparse bipartite graph shown in Fig.\ref{fig:2}(a). Similarly, the connections between variables $\bar{x}_{kl}$ and the weight-3 syndromes $\bar{s}_{klm}^{(3)}$ are depicted in Fig.\ref{fig:2}(b). The number of elements in $\bar{\boldsymbol{x}}$ is $N_{v}=\tbinom{K}{2}$, while the number of elements in $\boldsymbol{\bar{s}}^{(4)}(\bar{\boldsymbol{x}})$ and $\boldsymbol{\bar{s}}^{(3)}(\bar{\boldsymbol{x}})$ is $N_{c}=\tbinom{K-1}{2}$ and $N_{c}=\tbinom{K}{3}$, respectively. In the terminology of graph theory, the $N_{v}$ elements in $\bar{\boldsymbol{x}}$ constitute variable nodes (VNs) and the $N_{c}$ elements in $\boldsymbol{\bar{s}}^{(4)}(\bar{\boldsymbol{x}})$ or $\boldsymbol{\bar{s}}^{(3)}(\bar{\boldsymbol{x}})$ constitute check nodes (CNs). The matrix $\boldsymbol{H}$ or $\boldsymbol{H}'$ has $N_{c}$ rows and $N_{v}$ columns where each row $i$ represents a CN and each column $j$ represents a VN; If the entry $H_{ij}=1$, VN $j$ is connected to CN $i$ by an edge. Thus, each edge connecting VN and CN corresponds to an entry $1$ in the row and column of $\boldsymbol{H}$ or $\boldsymbol{H}'$. Let $d_{c}$ be the number of $1$s in each row and $d_{v}$ be the number of $1$s in each column of $\boldsymbol{H}$ or $\boldsymbol{H}'$. These numbers characterize the number of edges connected to a VN and a CN, respectively, called row and column weights. The matrix $\boldsymbol{H}'$ is regular, where the row weight $d_{c}=3$ is common for all the CNs and the column weight $d_{v}=K-2$ is common for all the VNs. In contrast, the matrix $\boldsymbol{H}$ is irregular, and its weights depend on the associated nodes, which are at most 4. Since the weight-3 syndrome is written as a linear combination of three elements in $\bar{\boldsymbol{x}}$: $\bar{s}_{klm}^{(3)}=\bar{x}_{kl}\oplus\bar{x}_{lm}\oplus\bar{x}_{km}$, where $1\leq k<l<m\leq K$, its spin representation $s_{klm}^{(3)}$ is written as a product of three elements in $\boldsymbol{x}$: $s_{klm}^{(3)}=x_{kl}x_{lm}x_{km}$ (see Eq.(\ref{eq:8})).  Similarly, $s_{klmn}^{(4)}$  is written as a product of four elements in $\boldsymbol{x}$ if $\boldsymbol{x}$ is complemented by the fictitious spin variables  $x_{ii} \,(i=2,\ldots,K-1)$ having fixed value 1: $s_{klmn}^{(4)}=x_{km}x_{lm}x_{ln}x_{kn}$ \cite{lechnerQuantumAnnealingArchitecture2015}. It should be noted that any element in $\boldsymbol{\bar{s}}^{(4)}(\bar{\boldsymbol{x}})$ can be written as a linear combination of appropriate elements in $\boldsymbol{\bar{s}}^{(3)}(\bar{\boldsymbol{x}})$, and vice versa. Similarly, in the spin representation, any element in $\boldsymbol{s}^{(4)}(\boldsymbol{x})$ can be written as a product of appropriate elements in $\boldsymbol{s}^{(3)}(\boldsymbol{x})$, and vice versa.

\begin{table}[tb]
\begin{tabular}{|c|c|c|c|c|}
\hline 
$K$ & $\tbinom{K}{2}$ & $\tbinom{K-1}{2}$ & $\tbinom{K}{3}$ & $K-2$\tabularnewline
\hline 
\hline 
4 & 6 & 3 & 4 & 2\tabularnewline
\hline 
5 & 10 & 6 & 10 & 3\tabularnewline
\hline 
6 & 15 & 10 & 20 & 4\tabularnewline
\hline 
7 & 21 & 15 & 35 & 5\tabularnewline
\hline 
\end{tabular}\caption{Parameters related to Fig.\ref{fig:2} for $4\leq K\leq7$.}
\end{table}

\begin{figure}[tb]
\includegraphics[viewport=340bp 230bp 580bp 390bp,clip,scale=0.6]{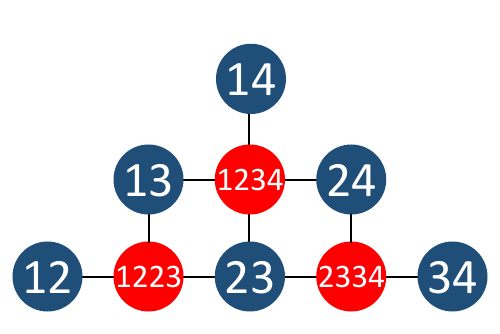}\caption{Bipartite graph topologically equivalent to Fig.\ref{fig:2}(a). This graph avoids any edge crossings. 
\label{fig:3}}
\end{figure}

Let us consider the graph in Fig.\ref{fig:3}, which is topologically equivalent to Fig.\ref{fig:2}(a). We can see that this is nothing but the SLHZ system. It can also be seen that $H^{code}\left(\boldsymbol{z}\right)$ in Eq.(\ref{eq:9}) agrees with the Hamiltonian of the SLHZ system. Therefore, it was confirmed that there is a close connection between the SLHZ system and the word MAP decoding of the LDPC codes. Furthermore, since the two formulations for word MAP decoding, i.e. those based on $H^{code}\left(\boldsymbol{z}\right)$ (Eq.(\ref{eq:9})) and those based on $H^{source}\left(\boldsymbol{Z}\right)$ (Eq.(\ref{eq:5})) are mathematically equivalent, it follows that finding the ground state of the Hamiltonian of the SLHZ system is equivalent to finding the ground state of the spin glass. 

\subsection{Decoding readout of the SLHZ system }

\subsubsection*{Bit flipping decoding algorithm}

Now, let us consider decoding the readout of the SLHZ system from the perspective of the LDPC codes. To begin with, we consider the simplest model, i.e. the correction of errors caused by i.i.d. noise. This model is reasonable for noisy communication systems and QA when measurement errors dominate over other sources of error in the readout. We show that a simple decoding algorithm with very low decoding complexity can be used in this case. Our decoding algorithm only uses information about the syndrome vector $\boldsymbol{s}(\boldsymbol{x})$  to eliminate errors in the current $\boldsymbol{x}$. Furthermore, our algorithm is a symmetric decoder since all spin variables are treated symmetrically. It should be noted here that the syndrome vector depends only on the error pattern $\boldsymbol{e}=\boldsymbol{x}\circ\boldsymbol{z}=(e_{12},\ldots,e_{K-1\,K})\in\left\{ \pm1\right\} ^{\tbinom{K}{2}}$, where $\circ$ denotes componentwise multiplication. This is because 
\begin{equation}
\boldsymbol{s}(\boldsymbol{z})=\left(\prod_{\left\{ j:H_{1j}=1\right\} }z_{j},\ldots,\prod_{\left\{ j:H_{N_{c}j}=1\right\} }z_{j}\right)=(+1,\ldots,+1)
\end{equation}
holds for any code-state $\boldsymbol{z}$, and therefore 
\begin{equation}
\boldsymbol{s}(\boldsymbol{x})=\boldsymbol{s}(\boldsymbol{z}\boldsymbol{e})=\boldsymbol{s}(\boldsymbol{z})\circ\boldsymbol{s}(\boldsymbol{e})=\boldsymbol{s}(\boldsymbol{e})
\end{equation}
holds. Note that, in general, we only know $\boldsymbol{x}$ but never know $\boldsymbol{e}$. Since LDPC codes are linear codes, and the AWGN channel is assumed to be symmetric, i.e. $P(y_{i}|z_{i})=P(-y_{i}|-z_{i})$ holds, the success probability of decoding is independent of the input code-state $\boldsymbol{z}$ \cite{richardsonCapacityLowdensityParitycheck2001b}.

\begin{figure*}
\includegraphics[viewport=150bp 160bp 800bp 400bp,clip,scale=0.65]{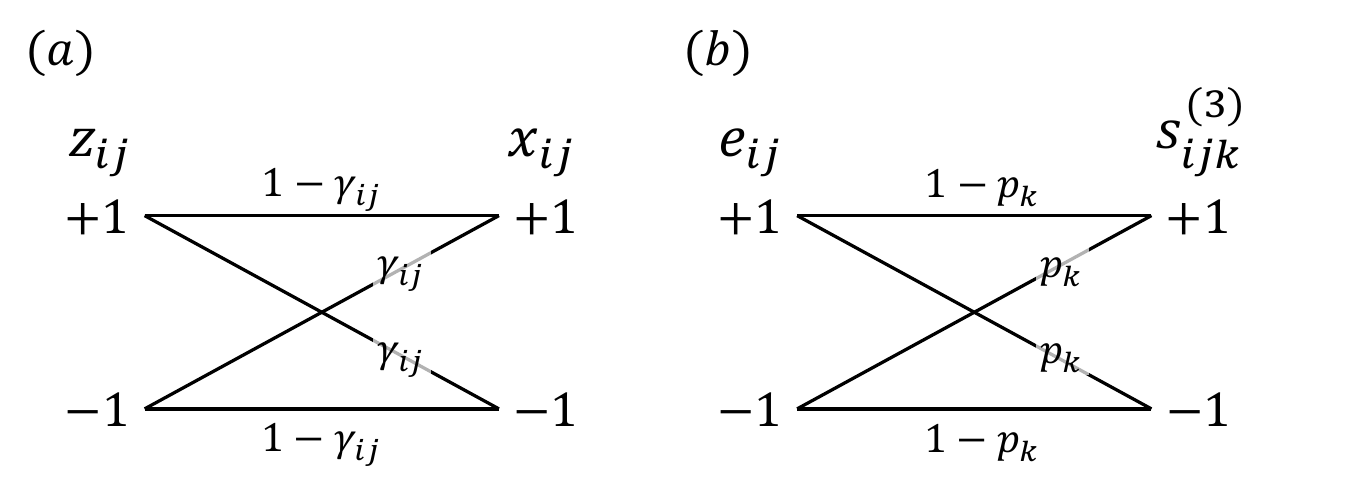}
\caption{Binary symmetric channel associated with the parameters $\gamma_{ij}$ and $p_{k}$.
\label{fig:4}}
\end{figure*}

The proposed decoding algorithm corresponds to Gallager's BF decoding algorithm in the spin representation \cite{gallagerLowDensityParityCheckCodes1962,gallagerLowDensityParityCheckCodes1963}. It is based on Massey's APP (a posteriori probability) threshold decoding algorithm \cite{masseyThresholdDecoding1962}. The weight-3 syndrome $s_{ijk}^{(3)}\left(\boldsymbol{x}\right)=x_{ij}x_{jk}x_{ik}=e_{ij}e_{jk}e_{ik}$ rather than the weight-4 syndrome is used to decode the current decision $\boldsymbol{x}$ because it has the advantage of treating all variables (VN and CN) symmetrically \cite{pastawskiErrorCorrectionEncoded2016}. The trade-off is that more constraints may increase the decoding complexity. However, as will be shown later, this is not the case for us since we can take advantage of symmetry to reduce computational costs. To obtain the best estimate $e_{ij}^{*}\in\left\{ \pm1\right\} $ of error $e_{ij}$, we use the syndromes $s_{ijk}^{(3)}\left(\boldsymbol{x}\right)$ with $k\neq i,j$, which are considered an $K-2$ parity checks orthogonal on $e_{ij}$ \cite{masseyThresholdDecoding1962}. The value of the syndrome $s_{ijk}^{(3)}\left(\boldsymbol{x}\right)$ indicates whether the parity-check equation is satisfied (being $+1$) or violated (being $-1$). Based on this value, we decide whether a spin should be flipped. There are $N_{v}=\tbinom{K}{2}$ choices for the set $\left\{ i,j\right\} $, and the associated syndromes $s_{ijk}^{(3)}\left(\boldsymbol{x}\right)$'s are computed from a hard decision $\boldsymbol{x}$ of the observation $\boldsymbol{y}$. The estimate $e_{ij}^{*}$ is obtained from the APP decoding by weighted majority voting \cite{masseyThresholdDecoding1962} 
\begin{equation}
e_{ij}^{*}=\mathrm{sign}\left(w_{0}+\sum_{k\neq i,j}^{K}w_{k}s_{ijk}^{(3)}\left(\boldsymbol{x}\right)\right)=\mathrm{sign}\left[\varDelta_{ij}\left(\boldsymbol{x}\right)\right],\label{eq:21}
\end{equation}
where the positive parameter $w_{0}$ is the weighting factor associated with the reliability $\left|\theta_{ij}\right|$ of the hard decision $x_{ij}=\mathrm{sgn}\left[y_{ij}\right]$:
\begin{equation}
w_{0}=\left|\theta_{ij}\right|=\log\frac{1-\gamma_{ij}}{\gamma_{ij}}.
\end{equation}
Here, $w_{0}=\beta\left|y_{ij}\right|$ for the AWGN channel and $\gamma_{ij}$ is the error probability that $x_{ij}$ is in error 
\begin{equation}
\gamma_{ij}=P\left(x_{ij}=-z_{ij}\right)=P\left(e_{ij}=-1\right).
\end{equation}
The positive parameter $w_{k}$ is a weighting factor associated with the reliability of the $k$th parity check for the decision of $e_{ij}$:
\begin{equation}
w_{k}=\log\frac{1-p_{k}}{p_{k}},
\end{equation}
where $p_{k}$ is the probability that a decision based on $s_{ijk}^{(3)}$ is in error: 
\begin{equation}
p_{k}=P\left(s_{ijk}^{(3)}=-e_{ij}\right)=P\left(e_{jk}e_{ik}=-1\right).
\end{equation}
In Eq.(\ref{eq:21}), $e_{ij}^{*}=-1$ means that we should invert the sign of $x_{ij}\rightarrow x_{ij}e_{ij}=-x_{ij}$. Function $\varDelta_{ij}\left(\boldsymbol{x}\right)$ is called the inversion function and determines whether to invert the sign of the spin variable $x_{ij}$  ($e_{ij}^{*}=-1$) or not ($e_{ij}^{*}=1$)  \cite{wadayamaGradientDescentBit2010,sundararajanNoisyGradientDescent2014a}. Parameters $\gamma_{ij}$ and $p_{k}$ are considered to be crosstalk parameters that characterize the binary symmetric channels shown in Fig.\ref{fig:4}. We should note that the probability $p_{k}$ is given by the probability of an odd number of $-1$s among the errors exclusive of $e_{ij}$ that are checked by $s_{ijk}^{(3)}\left(\boldsymbol{x}\right)=e_{ij}e_{jk}e_{ik}$ so that it is given by \cite{masseyThresholdDecoding1962} 
\begin{equation}
p_{k}=P\left(e_{jk}e_{ik}=-1\right)=\frac{1}{2}\left(1-\left(1-2\gamma_{jk}\right)\left(1-2\gamma_{ik}\right)\right).
\end{equation}

\begin{figure}[tb]
\includegraphics[viewport=290bp 160bp 650bp 390bp,clip,scale=0.7]{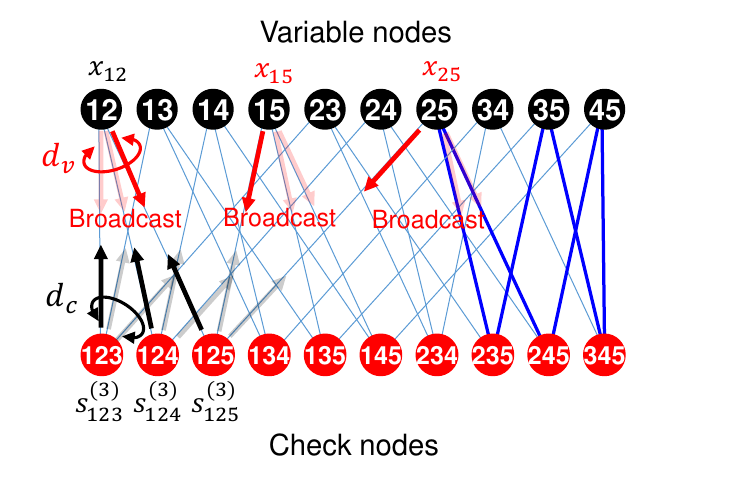}
\caption{Bipartite graph for $K=5$ logical spins. Solid blue lines show an example of the shortest loop of edges connecting VNs and CNs, which has a length of 6.
\label{fig:5}}
\end{figure}

Since we assumed i.i.d. noise, $\gamma_{ij}=\gamma_{0}$ holds for every $\left\{ i,j\right\} $ for which $\gamma_{0}$ is a constant. In this case, assuming $1\leq\gamma_{0}<\tfrac{1}{2}$, we can confirm that $w_{0}$ can be approximated as $w_{0}=w_{k}$ for any $k$. Thus, we obtain the following algorithm: we calculate the best estimate by 
\begin{equation}
e_{ij}^{*}=\mathrm{sign}\left(1+\sum_{k\neq i,j}^{K}s_{ijk}^{(3)}\left(\boldsymbol{x}\right)\right).\label{eq:27}
\end{equation}
This equation means that if the majority vote of $\left\{ 1,s_{ij1}^{(3)}\left(\boldsymbol{x}\right),\ldots,s_{ijK}^{(3)}\left(\boldsymbol{x}\right)\right\} \in\left\{ \pm1\right\} ^{K+1}$ is negative, $x_{ij}$ should be flipped to increase the value of $\sum_{k\neq i,j}^{K}s_{ijk}^{(3)}\left(\boldsymbol{x}\right)$. Thus, Eq.(\ref{eq:27}) is reduced to the following equation for the best estimate $z_{ij}\in\left\{ \pm1\right\} $: 
\begin{equation}
z_{ij}=x_{ij}e_{ij}^{*}=\mathrm{sign}\left(x_{ij}+\sum_{k\neq i,j}^{K}x_{jk}x_{ki}\right),\label{eq:28}
\end{equation}
where we used the fact $x_{ij}\in\left\{ \pm1\right\} $. Note that this expression is represented only in terms of the values of VNs. Eq.(\ref{eq:27})  can be interpreted as the Gallager's BF decoding \cite{gallagerLowDensityParityCheckCodes1962,gallagerLowDensityParityCheckCodes1963}.
Fig.\ref{fig:5} shows the relevant bipartite graph for $K=5$. Note that the graph associated with the SLHZ system is more loopy, with a minimum length of 4, whereas this graph is less loopy, with a minimum length of 6. The initial decision $\boldsymbol{x}$ on the VNs, hard-decided by $\boldsymbol{y}$, is broadcasted to its adjacent CNs connected by edges. Each CN then reports the syndrome value to its adjacent VNs connected by the edge. All VNs update their values simultaneously according to Eq.(\ref{eq:27}), representing a majority vote of 1 and the associated value of adjacent CNs.

\subsubsection*{Application to the SLHZ system}

Next, let us describe how our decoding applies to the SLHZ system. Consider a matrix representation of the state of the SLHZ system. Let us introduce the $K\times K$ symmetrized matrix $\hat{\boldsymbol{z}}$ whose elements are given by spin variable $z_{ij}=z_{ji}\in\left\{ \pm1\right\}$ and have unit diagonal elements, namely, 
\begin{eqnarray}
\hat{\boldsymbol{z}}
&=&\left[\begin{array}{cccccc}
1 & z_{12} & z_{13} & \cdots & z_{1\,K-1} & z_{1\,K}\\
z_{21} & 1 & z_{23} & \cdots & z_{2\,K-1} & z_{2\,K}\\
z_{31} & z_{32} & 1 & \cdots & z_{3\,K-1} & z_{3\,K}\\
\vdots & \vdots & \vdots & \ddots & \cdots & \vdots\\
z_{K-1\,1} & z_{K-1\,2} & z_{K-1\,3} & \cdots & 1 & z_{K-1\,K}\\
z_{K\,1} & z_{K\,2} & z_{K\,3} & \cdots & z_{K\,K-1} & 1
\end{array}\right]\nonumber\\
&\in&\left\{\pm1\right\} ^{K\times K}.
\end{eqnarray}
We use similar notations $\hat{\boldsymbol{x}}$ and $\hat{\boldsymbol{e}}$  for the symmetrized matrices associated with current $\boldsymbol{x}$ and error pattern $\boldsymbol{e}$. They satisfy $\hat{\boldsymbol{x}}=\hat{\boldsymbol{z}}\circ\hat{\boldsymbol{e}}$. Hereafter, we will call $\hat{\boldsymbol{e}}$ the error matrix. Then, Eq.(\ref{eq:28}) can be conveniently written as 
\begin{equation}
\hat{\boldsymbol{z}}=\mathcal{F}\left(\hat{\boldsymbol{x}}\right)=\mathrm{sign}\left[\hat{\boldsymbol{x}}\left(\hat{\boldsymbol{x}}-\boldsymbol{I}_{K\times K}\right)\right],\label{eq:30}
\end{equation}
where we assume the sign function is componentwise in this equation. This operation updates the $\tbinom{K}{2}$ elements in $\hat{\boldsymbol{x}}$ in parallel. Therefore, Eq.(\ref{eq:30}) represents a parallel BF algorithm. We further consider the iterative operation of $\mathcal{F}$, i.e., 
\begin{equation}
\hat{\boldsymbol{z}}^{(n)}=\hat{\boldsymbol{z}}\circ\hat{\boldsymbol{e}}^{(n)}=\mathcal{F}^{(n)}\left(\hat{\boldsymbol{x}}\right).\label{eq:31}
\end{equation}
If $\hat{\boldsymbol{z}}^{(n)}\rightarrow\hat{\boldsymbol{z}}$, or in other words, $\hat{\boldsymbol{e}}^{(n)}\rightarrow\hat{\boldsymbol{1}}_{K\times K}$ at $n=n_{0}$, where $\hat{\boldsymbol{1}}_{K\times K}$ is $K\times K$ matrix whose entries are all one (all-one matrix), we say that the decoding was succeeded after $n_{0}$ iteration of the BF decoding. In the case of success, it follows that $s_{ijk}^{(3)}\left(\hat{\boldsymbol{z}}\right)=1$ for a set of possible $\left\{ i,j,k\right\} $ at $n=n_{0}$. Our algorithm can safely be stated as an iterative process that gradually increases the syndrome values by flipping spins according to the majority vote of an associated set of relevant syndrome values.

\section{Experimental demonstration\label{sec:4}}

We demonstrate the validity and performance of the BF algorithm and compare it to those of the standard BP algorithm and word MAP decoding using Monte Carlo sampling. In addition, the results are shown when the BF decoding is applied to a stochastically sampled readout of the SLHZ system.

\subsection{I.i.d. noise model}

\begin{figure*}
\includegraphics[viewport=150bp 20bp 800bp 500bp,clip,scale=0.75]{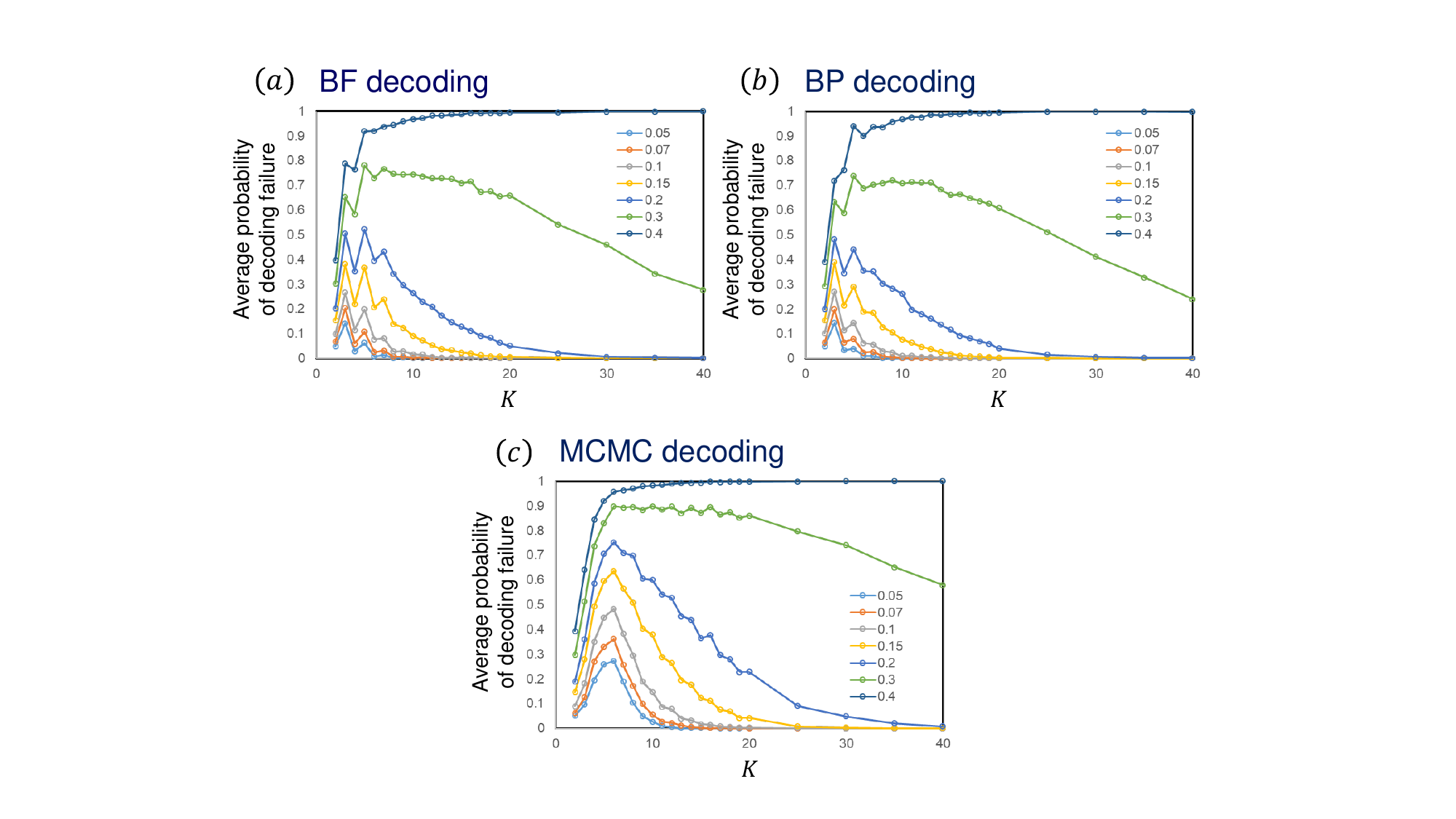}

\caption{Comparison of performance of (a) BF, (b) BP, and (c) MCMC decodings. Assuming that the error probability $\varepsilon$ is common to all physical spins, the average probabilities of decoding failure are plotted as a function of the number $K$ of logical spins (associated with $\tbinom{K}{2}$ physical spins) for seven values $\varepsilon$. Each data point was obtained by averaging over 5000 error matrix realizations. The BF and BP algorithms were iterated five times for each realization, and the MCMC sampling was iterated $\tbinom{K}{2}$ times. We considered a tie in the majority voting as a failure in the BF decoding, although we can resolve a tie by introducing a tie-break rule, such as coin tossing.
\label{fig:6}}
\end{figure*}

\begin{figure*}
\includegraphics[viewport=100bp 220bp 880bp 480bp,clip,scale=0.65]{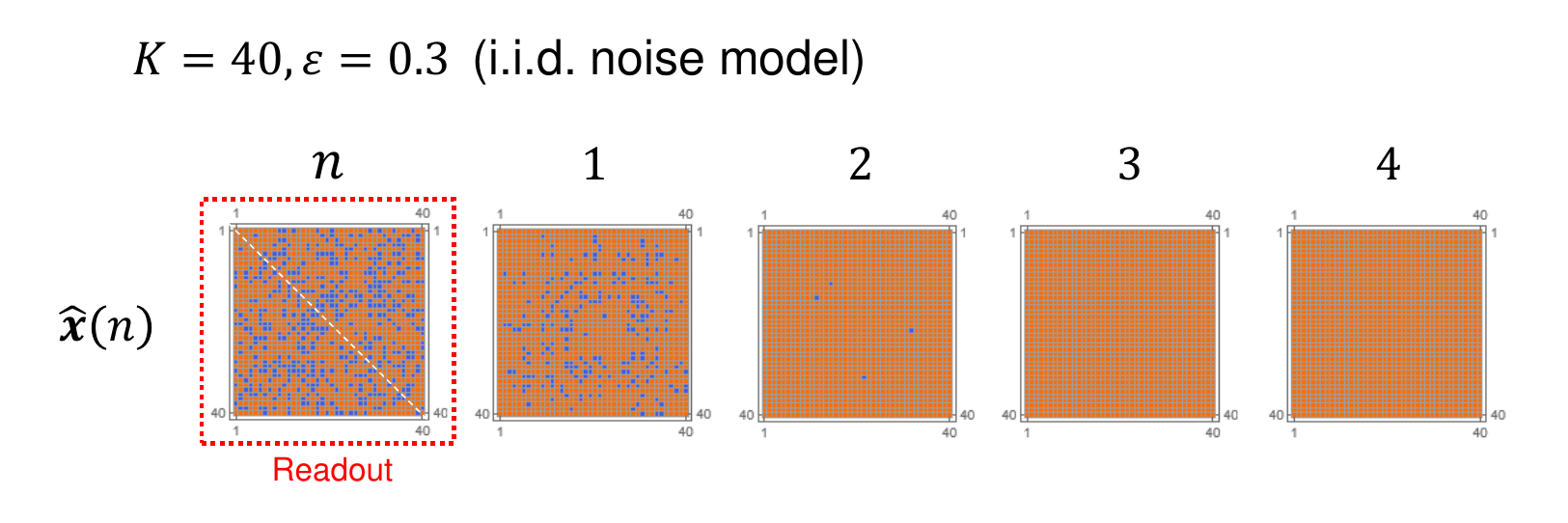}
\caption{An example of successfully decoded results by the BF decoding when $K=40$ and $\varepsilon=0.3$. The initial readout was generated in accordance with the i.i.d. noise model. The estimated matrix $\hat{\boldsymbol{x}}$ is plotted from left to right in increasing order of iterations $n$. The blue pixels represent spins with error.
\label{fig:7}}
\end{figure*}

First, the performance of the BF algorithm was investigated assuming i.i.d noise, a simple model of the noisy readout of the SLHZ system. PP previously studied this model with the BP algorithm \cite{pastawskiErrorCorrectionEncoded2016}. Note that the BF decoding in Eq.(\ref{eq:27}) depends only on the weight-3 syndromes, which treat all variables (VN and CN) symmetrically \cite{pastawskiErrorCorrectionEncoded2016}. In this case, without loss of generality, the performance of decoding can be analyzed using the assumption that all-one code-state $\boldsymbol{z}=\left(+1,\ldots,+1\right)$ has been transmitted \cite{richardsonCapacityLowdensityParitycheck2001b,vuffrayCavityMethodCoding}. In physics, this assumption corresponds to choosing the ferromagnetic gauge and representing the current decision $\boldsymbol{x}$ by an error pattern $\boldsymbol{e}$. We assume the input is all-one code-state, that is, $\hat{\boldsymbol{z}}=\hat{\boldsymbol{1}}_{K\times K}$. All the demonstrations were performed using the Mathematica$^{\circledR}$ Ver.14 platform on the Windows 11 operating system. We generated 5000 symmetric matrices $\hat{\boldsymbol{x}}=\hat{\boldsymbol{e}}$ with unit diagonal elements and other elements randomly assigned to $-1$ with probability $\varepsilon<\tfrac{1}{2}$ and $+1$ otherwise. After $n=5$ iterations, the BF decoding was checked to see if the error was corrected. Figure \ref{fig:6}(a) shows the performance of the BF algorithm, plotting the probability of decoding failure as a function of the number $K$ of logical spins for $K$ ranging from $2$ to $40$ and seven values of common bit error rate $\varepsilon$ $\left(=0.05,0.07,0.1,0.15,0.2,0.3,0.4\right)$. Note that the associated SLHZ model consists of $\tbinom{K}{2}$ physical spins. The failure probability falls steeply as $N$ increases if $\varepsilon$ is not too close to the threshold value $1/2$. A similar performance calculation was reproduced for the decoding based on the BP algorithm given by PP, assuming that it is iterated five times, as shown in Fig.\ref{fig:6}(b) \cite{pastawskiErrorCorrectionEncoded2016}. Comparing these figures, we can see that the performance of the BF algorithm is comparable to that of the BP algorithm. Note that the BP algorithm updates all marginal probabilities $P\left(x_{i}|\boldsymbol{y}\right)$ associated with the $\tbinom{K}{2}$ spins sequentially by passing a real-valued message between the associated VNs and CNs per a single iteration. In contrast, in the BF algorithm, all the $\tbinom{K}{2}$ variables $x_{ij}$ in the matrix $\hat{\boldsymbol{x}}$ are updated in parallel per a single iteration. Thus, in these performance evaluations, each variable was updated the same number of times, i.e., five times, in both the BF and BP algorithms. 
Fig.\ref{fig:7} is an example of a successfully decoded result when $K=40$ and $\varepsilon=0.3$. In this figure, each entry of the error matrix $\hat{\boldsymbol{e}}$ is plotted after $n=1,\ldots,4$ iterations of the operation in Eq.(\ref{eq:30}). The blue pixels correspond to spins with error, the number of which gradually decreases as rounds
of iteration are added. In this example, an error-free matrix was obtained after $n=3$ iterations. Such results were observed for more than $70$ \% of the error matrices generated. 

In addition to the BF and BP algorithms, let us focus on the word MAP decoding introduced in Sec.\ref{subsec:2-B}  and compare its performance with the above algorithms. The word MAP decoding says that the code-state is the ground state of the Hamiltonian, as shown in Eq.(\ref{eq:9}),  
\begin{equation}
H^{code}(\hat{\boldsymbol{x}})\equiv\gamma\sum_{\{ i,j,k\} }\frac{1-s_{ijk}^{(3)}(\hat{\boldsymbol{x}})}{2},\label{eq:32}
\end{equation}where $\hat{\boldsymbol{x}}\in\{ \pm1\} ^{K\times K}$ is the matrix representing the general state of the SLHZ system. In general, $H^{code}(\hat{\boldsymbol{x}})\geq0$ and $H^{code}(\hat{\boldsymbol{x}})=0$ if and only if $\hat{\boldsymbol{x}}$ is a code-state. In Eq.(\ref{eq:32}), the correlation term (the first term in Eq.(\ref{eq:9})) was omitted. This is because the all-one code-state assumption uses no information about the observation $\boldsymbol{y}$. Under this assumption, the ground state of $H^{code}(\hat{\boldsymbol{x}})$ is given by 
\begin{equation}
\hat{\boldsymbol{z}}=\underset{\hat{\boldsymbol{x}}\in\{ \pm1\} ^{K\times K}}{\arg\min}H^{code}(\hat{\boldsymbol{x}})=\hat{\boldsymbol{1}}_{K\times K}.\label{eq:33}
\end{equation}
The performance of word MAP decoding was evaluated based on the Hamiltonian $H^{code}(\hat{\boldsymbol{x}})$ given by Eq.(\ref{eq:32}). We generated 5000 random symmetric error matrices with bit error rate $\varepsilon$ as stated before. The classical MCMC sampler was used to search for the ground state of $H^{code}(\hat{\boldsymbol{x}})$ from the initial state $\hat{\boldsymbol{x}}=\hat{\boldsymbol{e}}$. The hyperparameter $\gamma$ was assigned $\gamma\approx1$, which was found to be experimentally optimal. We examined whether the ground state $\hat{\boldsymbol{z}}$ could be found in a set of estimates $\{ \hat{\boldsymbol{x}}\} $ sampled by MCMC. In this sampling, we used rejection-free MCMC sampling, in which all self-loop transitions are removed from the standard MCMC \cite{nambuRejectionFreeMonteCarlo2022,nambuErrorCorrectionParityencodingbased2024}, and the size of $\{ \hat{\boldsymbol{x}}\} $ was restricted to $\tbinom{K}{2}$. Figure \ref{fig:6}(c) indicates the performance of the word MAP decoding using the MCMC sampler, which we called the MCMC decoding. Although its performance is not as good as that of the BF and BP algorithms, it shows a similar dependence on $K$ and $\varepsilon$. This result is quite reasonable as well as suggestive. Later, we will discuss the reason.

\subsection{Beyond i.i.d. noise model}

As the development of QA devices is ongoing and classical simulation of a quantum-mechanical many-body system is computationally challenging, it is difficult to verify whether our BF decoding algorithm provides good protection against noise in physical spin readouts of QA, not only by classical computer simulations but also by an actual QA device. In this study, we tested the potential of our BF decoding algorithm by simulating the hard-decided readout of spins in the SLHZ system with the data stochastically sampled by a classical MCMC sampler.

We used the following Hamiltonian to obtain stochastically sampled data for the SLHZ system: 
\begin{equation}
H^{code}(\hat{\boldsymbol{x}})\equiv-\beta\sum_{\{ i,j\} }J_{ij}x_{ij}+\gamma\sum_{\{ i,j,k,l\} }\frac{1-s_{ijkl}^{(4)}(\hat{\boldsymbol{x}})}{2},\label{eq:34}
\end{equation}
where $J_{ij}\in\mathbb{R}$ is a coupling constant identified with the channel observation $y_{ij}\in\mathbb{R}$ in the AWGN channel model. Parameters $\left\{ \beta,\gamma\right\} $ are independent annealing parameters to be adjusted. If the weight-4 syndrome is defined by 
\begin{equation}
s_{ijkl}^{(4)}(\hat{\boldsymbol{x}})=x_{ik}x_{jk}x_{jl}x_{il},
\end{equation}
with the assumption $x_{ii}=1$ for $i=1,\ldots,K$, Eq.(\ref{eq:34}) is just the Hamiltonian of the SLHZ system given in Fig.\ref{fig:3}. The $(K-1)$-degenerated ground state of the second term in Eq.(\ref{eq:34}) defines the code-states. Sampling from the low-temperature equilibrium state $\hat{\boldsymbol{x}}$  of  $H^{code}(\hat{\boldsymbol{x}})$, we can see which is most likely to be the code-state given by the matrix $\hat{\boldsymbol{z}}=\boldsymbol{Z}^{T}\boldsymbol{Z}$ that minimizes $H^{code}(\hat{\boldsymbol{x}})$, as shown in Eq.(\ref{eq:33}), where $\boldsymbol{Z}=(Z_{1},\ldots,Z_{K})\in\{ \pm1\} ^{K}$ is the source-state. Therefore, optimization using this Hamiltonian corresponds to nothing but the word MAP decoding. However, in this case, $\hat{\boldsymbol{z}}=\hat{\boldsymbol{1}}_{K\times K}$ cannot be assumed in the evaluation of performance, in contrast to Eq.(\ref{eq:33}). This is because both the correlation term (first term) as well as the penalty term (second term) in Eq.(\ref{eq:34}) violate the symmetry conditions required for all-one code-state assumption to hold in this case. As a result, the performance of the word MAP decoding based on Eq.(\ref{eq:34}) depends not only on the error distribution $\hat{\boldsymbol{e}}$ but also on $\hat{\boldsymbol{z}}$ . 

\begin{figure*}
\includegraphics[viewport=100bp 100bp 840bp 440bp,clip,scale=0.65]{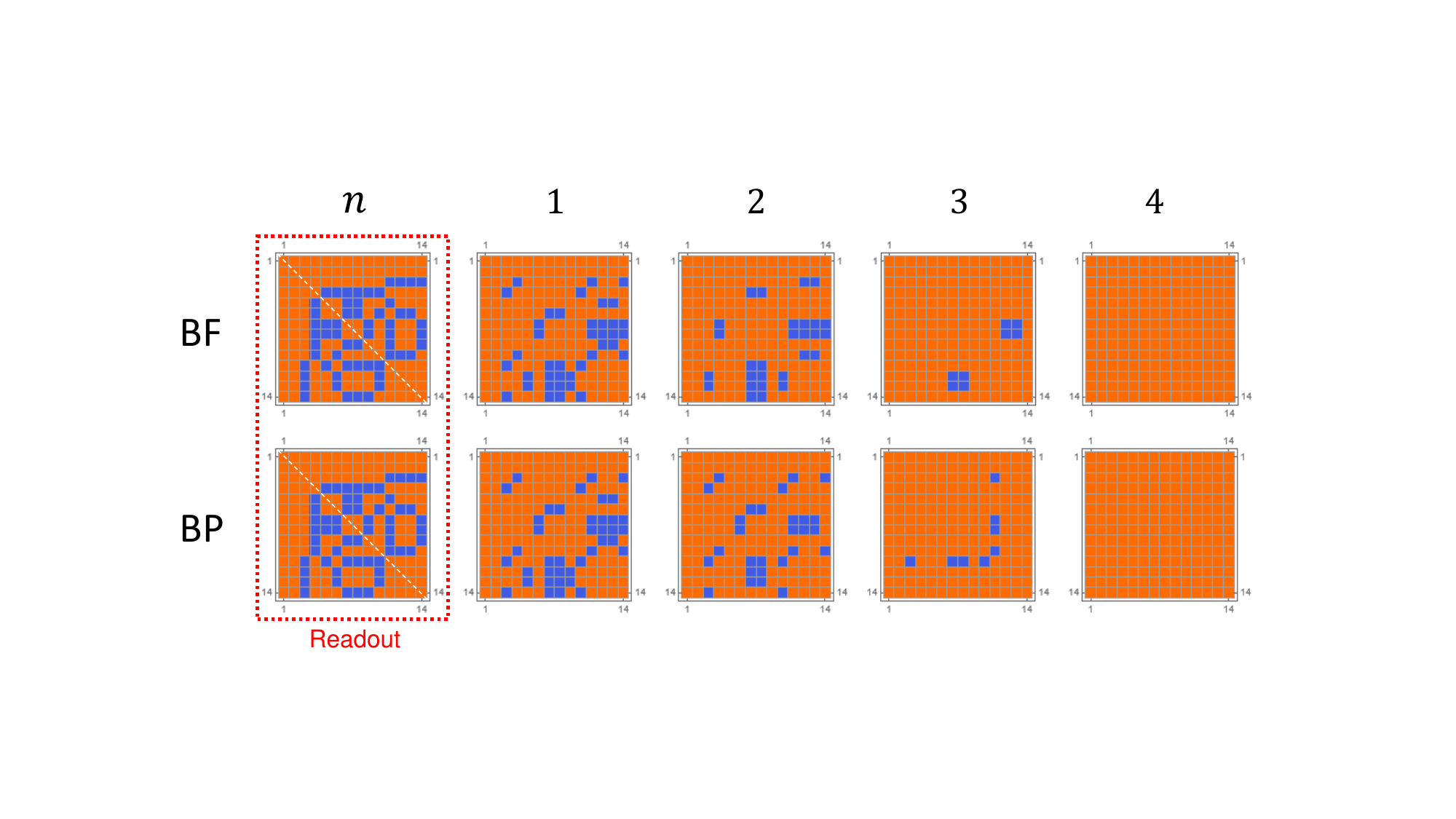}
\caption{Examples of successful BF and BP decoding for $K=14$ logical spins. An MCMC sampler sampled readout based on the Hamiltonian $H^{code}(\hat{\boldsymbol{x}})$ of Eq.(\ref{eq:34}). See details in the main text. The readout matrix $\hat{\boldsymbol{x}}$ is plotted from left to right in order of increasing number of iterations $n$ of the two algorithms. The blue pixels represent spins with error.
\label{fig:8}}
\end{figure*}

Our BF algorithm is also valid for the readout of the SLHZ system sampled by a classical MCMC sampler. We show an illustrative example to see this. We simulated readouts $\hat{\boldsymbol{x}}\in\{ \pm1\} ^{K\times K}$ of the SLHZ system associated with a spin glass problem for $K=14$  ($K_{14}$)  as a toy model using the MCMC sampler. To this end, we generated a set of $12$ logical random instances on complete graphs $K_{14}$ with couplings $J_{ij}\in[-\tfrac{1}{4},\tfrac{1}{4}]$ chosen uniformly at random and all logical local fields set to zero, where the code-state $\hat{\boldsymbol{z}}$  for a given $\boldsymbol{J}=(J_{12},\ldots,J_{K-1\,K})\in\mathbb{R}^{\tbinom{K}{2}}$ was precomputed by brute force. The leftmost matrix plot in Fig.\ref{fig:8} visualizes an example of the error matrix $\hat{\boldsymbol{e}}=\hat{\boldsymbol{x}}\circ\hat{\boldsymbol{z}}$ associated with a sampled readout $\hat{\boldsymbol{x}}$ of the SLHZ system. In this example, the error distribution was different from the expected when assuming an i.i.d. noise model, as shown in Fig.\ref{fig:7}. This suggests that our BF algorithm, as well as the BP algorithm, can correct errors in the stochastically sampled readouts of the SLHZ system. Here, although we can not know the error rate for each spin a priori because the correct ground state is generally unknown a priori, a uniform i.i.d. error rate of 0.25 was assumed when running the BP algorithm.

The MCMC sampling was performed as follows: We sampled a sequence $\{ \hat{\boldsymbol{x}}\}$ of estimated states using a rejection-free MCMC sampler starting from a random state. After a certain number of MCMC samplings, we checked whether $\{ \hat{\boldsymbol{x}}\}$ involved $\hat{\boldsymbol{z}}$. If $\{ \hat{\boldsymbol{x}}\}$ involved $\hat{\boldsymbol{z}}$, we decided that the decoding was successful. Otherwise, it failed. The performance can be evaluated by repeating the same experiment independently, obtaining many sampled sequences $\{ \hat{\boldsymbol{x}}\}$, and measuring the average probability of obtaining a successfully decoded sequence. 

We call a search of the ground state $H^{code}(\hat{\boldsymbol{x}})$ using the MCMC sampler the MCMC decoding. In addition to this, we also performed two-stage hybrid decoding. In this strategy, the sequence $\{ \hat{\boldsymbol{x}}\}$ is sampled by the MCMC sampler in the first stage. Each element of the sequence $\{ \hat{\boldsymbol{x}}\}$ is then decoded using the BF decoding in the second stage to correct the errors in each $\hat{\boldsymbol{x}}$ and updated to $\{ \hat{\boldsymbol{w}}\} $. Then, if $\{ \hat{\boldsymbol{w}}\} $ involved $\hat{\boldsymbol{z}}$, we decided that the decoding was successful. Otherwise, it failed. We call this strategy the MCMC-BF hybrid decoding. We can evaluate its performance as well. The MCMC decoding can be identified with simulated annealing. Similarly, the MCMC-BF hybrid decoding can be identified with simulated annealing and subsequent classical error correction. 

\begin{figure*}
\includegraphics[viewport=120bp 40bp 825bp 500bp,clip,scale=0.7]{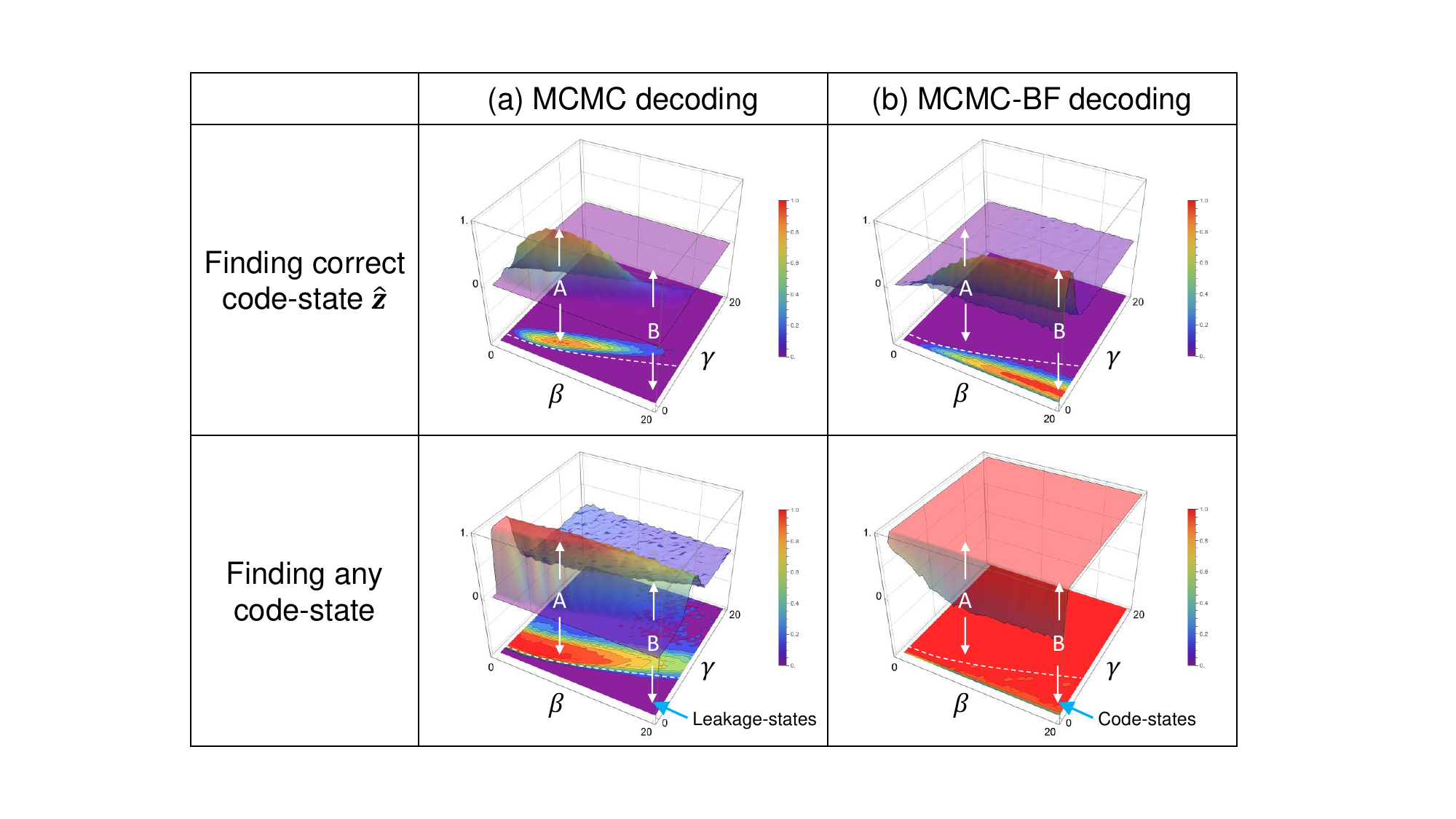}
\caption{Landscapes of the probability distribution for successful decoding. They are plotted as functions of the annealing parameters $\{ \beta,\gamma\} $. The left and right columns are the results of (a) MCMC decoding and (b) MCMC-BF hybrid decoding, respectively. On the other hand, the states searched in the upper and lower rows are different: they are the correct target state $\hat{\boldsymbol{z}}$ for the upper row and any code-states for the lower row. 
\label{fig:9}}
\end{figure*}

The performance of these decoding strategies depends not only on the annealing parameters $\left\{ \beta,\gamma\right\}$ but also on the strategy as shown in Fig.\ref{fig:9}. The columns (a) and (b) indicate landscapes of the success probability for (a) the MCMC decoding and (b) the MCMC-BF hybrid decoding, respectively. In these figures, the top two figures are the probability of success in finding the correct code-state $\hat{\boldsymbol{z}}=\boldsymbol{Z}^{T}\boldsymbol{Z}$, and the bottom two figures are the probability of finding any code-state $\hat{\boldsymbol{x}}=\boldsymbol{X}^{T}\boldsymbol{X}$, where $\boldsymbol{X}\in\{ \pm1\} ^{K}$ is any logical state. It is very important to note that the size of the sample $\{ \hat{\boldsymbol{x}}\} $ differs significantly between the two decoding strategies; it was $1200\tbinom{K}{2}$ for the MCMC decoding and $4\tbinom{K}{2}$ for the MCMC-BF hybrid decoding, which were determined by the number of iterations of the rejection-free MCMC loops. The readers can understand that the MCMC-BF hybrid decoding provides a better trade-off between error performance and decoding complexity, assuming that the decoding complexity of the BF decoding is negligible. We will revisit the validity of this assumption later. 

\begin{figure*}
\includegraphics[viewport=120bp 160bp 700bp 450bp,clip,scale=0.75]{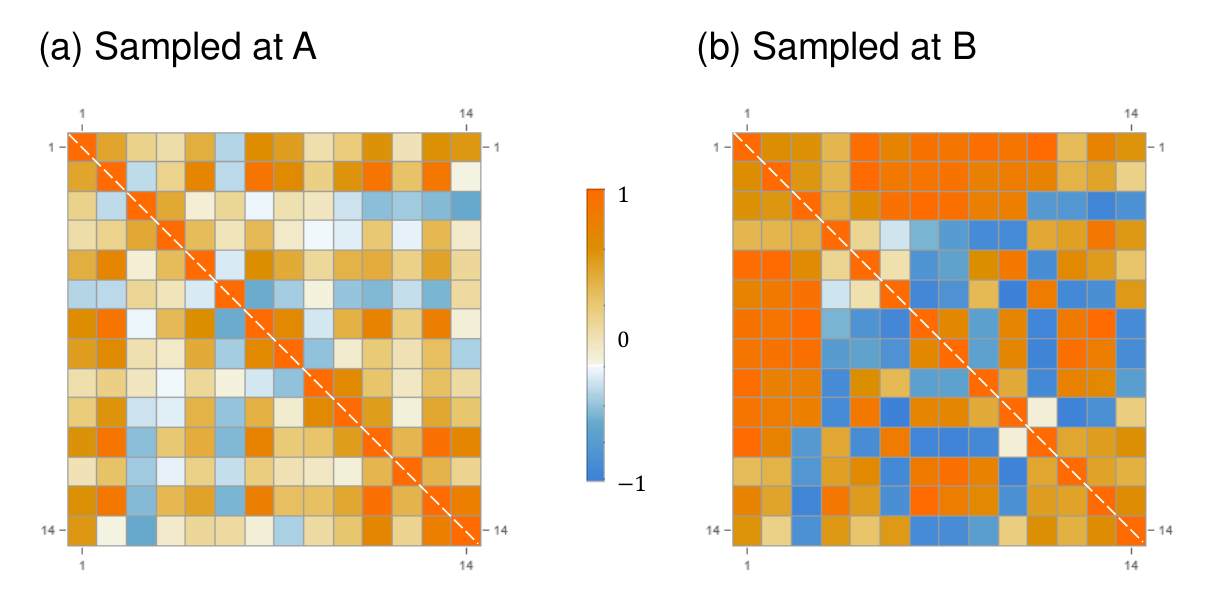}
\caption{Matrix plots representing the averaged error matrix $\langle \hat{\boldsymbol{e}}\rangle =\langle \hat{\boldsymbol{x}}\rangle \circ\hat{\boldsymbol{z}}$ after the MCMC sampling (but before the BF decoding), where each entry reflects the marginal error probability in inferring the correct sign of the associated entry. Warm (cold) colored pixels indicate that there is likely no error (error) in the inference of the corresponding spin. The left and right plots correspond to $\langle \hat{\boldsymbol{e}}\rangle$ when $\{ \hat{\boldsymbol{x}}\} $ was sampled with the parameter sets $A$ and $B$ in Fig.\ref{fig:9}, respectively. 
\label{fig:10}}
\end{figure*}

Let us note two arrows $A$ and $B$ denoted in Fig.\ref{fig:9}. These indicate two annealing parameter sets $A=\{ \beta_{A},\gamma_{A}\} $ and $B=\{ \beta_{B},\gamma_{B}\} $. The set $A$ corresponds to parameters for which the MCMC decoding is highly efficient, and the set $B$ corresponds to the parameters for which the MCMC-BF hybrid decoding is highly efficient. The left and right plots  in Fig. \ref{fig:10} are the matrix plots showing average error matrix $\langle \hat{\boldsymbol{e}}\rangle =\langle \hat{\boldsymbol{x}}\rangle \circ\hat{\boldsymbol{z}}$ over the set of samples $\{ \hat{\boldsymbol{x}}\} $ of the MCMC sampler when they were sampled with the parameter sets $A$ and $B$, respectively.  A positive (negative) entry in $\langle \hat{\boldsymbol{e}}\rangle $ indicates that there is likely no error (error) in the corresponding entry in $\hat{\boldsymbol{x}}$. Thus, each entry in $\langle \hat{\boldsymbol{e}}\rangle$ reflects a marginal error probability for inferring the correct sign for that entry. We can see that they depend on the annealing parameters. The average error matrix $\langle \hat{\boldsymbol{e}}\rangle$ in Fig.\ref{fig:10}(b) is consistent with the error matrix $\hat{\boldsymbol{e}}$ in Fig.\ref{fig:8}. In fact, the error matrix $\hat{\boldsymbol{e}}$ in Fig.\ref{fig:8} was an element of $\{\hat{\boldsymbol{x}}\}$ sampled by the MCMC sampling with the parameter set $B$. Note that the parameter set $B$ allowed little sampling for the code-state by the MCMC sampler in the first stage (see the lower left plot in Fig.\ref{fig:9}). This means that the state $\hat{\boldsymbol{x}}$ the MCMC sampler sampled in the first stage of the MCMC-BF decoding is not a code-state, which is quite reasonable because the subsequent BF decoding can correct errors in the state $\hat{\boldsymbol{x}}$ only if it is a leakage state. It is important to note that this is a feature of our BF decoding algorithm and is independent of the decoding algorithm in the first stage. Therefore, optimal annealing parameters $\{ \beta_{opt},\gamma_{opt}\} $ should differ between single and two-stage hybrid decodings. Our result suggests that optimal annealing parameters should be carefully studied when applying our BF decoding for readouts of QA; they might differ from the optimal parameters for QA when used alone. 

\section{Discussions\label{sec:5}}

\subsection{Relationship between MCMC-BF hybrid algorithm and other known BF algorithms\label{subsec:5-A}}

The MCMC-BF decoding algorithm includes MCMC decoding as a preprocessing step for the BF decoding algorithm. Both the MCMC and BF decoding algorithms are hard-decision algorithms. We consider the mutual relationship between these algorithms and similar existing hard-decision algorithms.

There are two obvious shortcomings in the postprocessing BF decoding algorithm. Introducing the MCMC decoding as a preprocessing step can solve these. The first one is that the BF algorithm neglects the soft information in the observation $\boldsymbol{y}$. As mentioned, the BF algorithm assumes a symmetric channel and treats all the spin variables (VNs) and syndromes (CNs) symmetrically, justifying the all-one code-state assumption. This is the result of neglecting the $\boldsymbol{y}$-dependent correlation term that breaks the necessary symmetry in the Hamiltonian $H^{code}(\hat{\boldsymbol{x}})$ in Eq.(\ref{eq:9}) (see also Eq.(\ref{eq:32})). Soft information should be considered to improve decoding, which is frequently considered by incorporating the reliability information, LLR $\theta(y_{i})$ , into the algorithm. In contrast, the BF decoding loses soft information since it approximates the weighted majority formula in Eq. (\ref{eq:21}) to the majority formula in Eq. (\ref{eq:27}) by assuming that $\gamma_{ij}=\gamma_{0}$ for the error probability of channel decision $x_{ij}$, which is reasonable in the i.i.d. noise model. In general, $\gamma_{ij}$ depends on the channel index $\left\{ i,j\right\} $. In the AWGN channel model, $\theta\left(y_{i}\right)=\beta y_{i}$ and Eq. (\ref{eq:13}), it follows that $\gamma_{ij}$ is small (large) if $\left|y_{ij}\right|$ is small (large) and that the hard decision $x_{ij}=\mathrm{sgn}\left[y_{ij}\right]$  is less (more) reliable.

Many researchers have developed sophisticated versions of the weighted BF (WBF). See details in Refs.\cite{khoaletrungNewDirectionLow,kennedymasundaThresholdBasedMultibit2017} and references therein. Alternatively, there is another approach to incorporate soft information. For example, Wadayama et al. proposed the Gradient Descent Bit Flipping (GDBF) algorithm \cite{wadayamaGradientDescentBit2010}. They are based on modifications to the inversion function, which may improve the decoding performance, although it increases decoding complexity. For example, the inversion functions employed in the BF, WBF, and GDBF algorithms are formally written as 
\begin{equation}
\varDelta_{k}^\mathrm{(BF)}(\boldsymbol{x})=1+\sum_{i\in M(k)}s_{i}(\boldsymbol{x}),\label{eq:36}
\end{equation}
\begin{equation}
\varDelta_{k}^\mathrm{(WBF)}(\boldsymbol{x})=\beta\left|J_{k}\right|+\sum_{i\in M(k)}w_{k}s_{i}(\boldsymbol{x}),\label{eq:37}
\end{equation}
and 
\begin{equation}
\varDelta_{k}^\mathrm{(GDBF)}(\boldsymbol{x})=J_{k}x_{k}+\sum_{i\in M(k)}s_{i}(\boldsymbol{x}),\label{eq:38}
\end{equation}
respectively. Here, $\boldsymbol{x}=(x_{1},\ldots,x_{N_{v}})\in\{ \pm1\} ^{N_{v}}$ the current decision (VNs), $\beta$ is a positive real parameter to be adjusted, and 
\begin{equation}
s_{i}(\boldsymbol{x})=\prod_{j\in N(i)}x_{j}\in\left\{ \pm1\right\}
\end{equation}
is $i$th syndrome (CN) for the decision $\boldsymbol{x}$ in the spin representation, $J_{k}$ is identified with channel observation $y_{k}$ in the AWGN channel model. Here, the set $N(i)=\left\{ j:H_{ij}=1\right\} $ is the VNs adjacent to an $i$th CN $\left(1\leq i\leq N_{c}\right)$ and the set $M(j)=\left\{ i:H_{ij}=1\right\} $ is the CNs adjacent to a $j$th VN $\left(1\leq j\leq N_{v}\right)$. Please refer to Fig.\ref{fig:2} for the definition of these sets. Note that as long as $s_{i}\left(\boldsymbol{x}\right)$ is a weight-3 syndrome, $\varDelta_{k}^\mathrm{(BF)}(\boldsymbol{x})$ is symmetric for the permutation of the elements of $\boldsymbol{x}$. In other words, $\varDelta_{k}^\mathrm{(WBF)}(\boldsymbol{x})$  is invariant under exchanging  $i\longleftrightarrow j$ for any pair of indices of VN other than $k$, justifying the all-one code-state assumption.
In contrast, weight-4 syndrome is not symmetric. Similarly, because $\varDelta_{k}^\mathrm{(WBF)}(\boldsymbol{x})$  involves weight $w_{k}$ in the second term and $\varDelta_{k}^\mathrm{(GDBF)}(\boldsymbol{x})$ involves $J_{k}$ in the first term, both depend on observation $\boldsymbol{y}$. Thus,  $\varDelta_{k}^\mathrm{(WBF)}(\boldsymbol{x})$ and $\varDelta_{k}^\mathrm{(GDBF)}(\boldsymbol{x})$ are not symmetric even if $s_{i}(\boldsymbol{x})$ is weight-3 syndrome. It is interesting to note that the inversion functions in Eq.(\ref{eq:36})-(\ref{eq:38}) can be formally derived from the following Hamiltonians: 
\begin{equation}
H^\mathrm{(BF)}(\boldsymbol{x})=-\sum_{i=1}^{N_{v}}x_{i}-\sum_{i=1}^{N_{c}}s_{i}(\boldsymbol{x}),
\end{equation}
\begin{equation}
H^\mathrm{(WBF)}(\boldsymbol{x})=-\beta\sum_{i=1}^{N_{v}}\left|J_{i}\right|x_{i}-\sum_{i=1}^{N_{c}}w_{i}s_{i}(\boldsymbol{x}),
\end{equation}
\begin{equation}
H^\mathrm{(GDBF)}(\boldsymbol{x})=-\frac{1}{2}\sum_{i=1}^{N_{v}}J_{i}x_{i}-\sum_{i=1}^{N_{c}}s_{i}(\boldsymbol{x}).
\end{equation}
If we note that when we invert the sign of $x_{k}$, that is, $x_{k}\rightarrow-x_{k}$, the increase in energy  $\Delta H_{k}^\mathrm{(X)}(\boldsymbol{x})$ is given by 
\begin{equation}
\Delta H_{k}^{(X)}(\boldsymbol{x})=2\varDelta_{k}^{(X)}(\boldsymbol{x}),
\end{equation}
where $X=\mathrm{BF}$, $\mathrm{WBF}$, or $\mathrm{GDBF}$. It follows that if $\varDelta_{k}^{(X)}(\boldsymbol{x})<0$, flipping the $k$th spin reduces the energy of the spin system. Therefore, the BF, WBF, and GDBF decoding algorithms can be considered deterministic algorithms that determine the most suitable spins to be flipped to reduce the energy $H_{k}^{(X)}(\boldsymbol{x})$ based on the inverse function $\varDelta_{k}^{(X)}(\boldsymbol{x})$.  In contrast to the BF algorithm, it should be noted that the BP algorithm essentially considers soft information since it is based on calculating a consistent marginal probability $P\left(x_{i}|\boldsymbol{y}\right)$ by exchanging real-valued messages between the VNs and CNs. On the other hand, the inversion function and the associated Hamiltonian for the MCMC decoding are formally written by 
\begin{equation}
\varDelta_{k}^\mathrm{(MCMC)}(\boldsymbol{x})=\beta J_{k}x_{k}+\frac{\gamma}{2}\sum_{i\in M(k)}s_{i}(\boldsymbol{x})
\end{equation}
and
\begin{equation}
H^\mathrm{(MCMC)}(\boldsymbol{x})=-\beta\sum_{i=1}^{N_{v}}J_{i}x_{i}+\gamma\sum_{i=1}^{N_{c}}\frac{1-s_{i}(\boldsymbol{x})}{2}.
\end{equation}
The MCMC decoding algorithm uses MCMC sampling to find $\boldsymbol{x}$ that reduces the energy $H^\mathrm{(MCMC)}(\boldsymbol{x})$ based on the inversion function $\varDelta_{k}^\mathrm{(MCMC)}(\boldsymbol{x})$. Therefore, the MCMC decoding at the first stage intrinsically includes the soft information in $J_{k}$ (namely $y_{k}$) in the correlation term, which resolves one shortcoming of the BF decoding.

\begin{figure*}
\includegraphics[viewport=80bp 30bp 840bp 500bp,clip,scale=0.5]{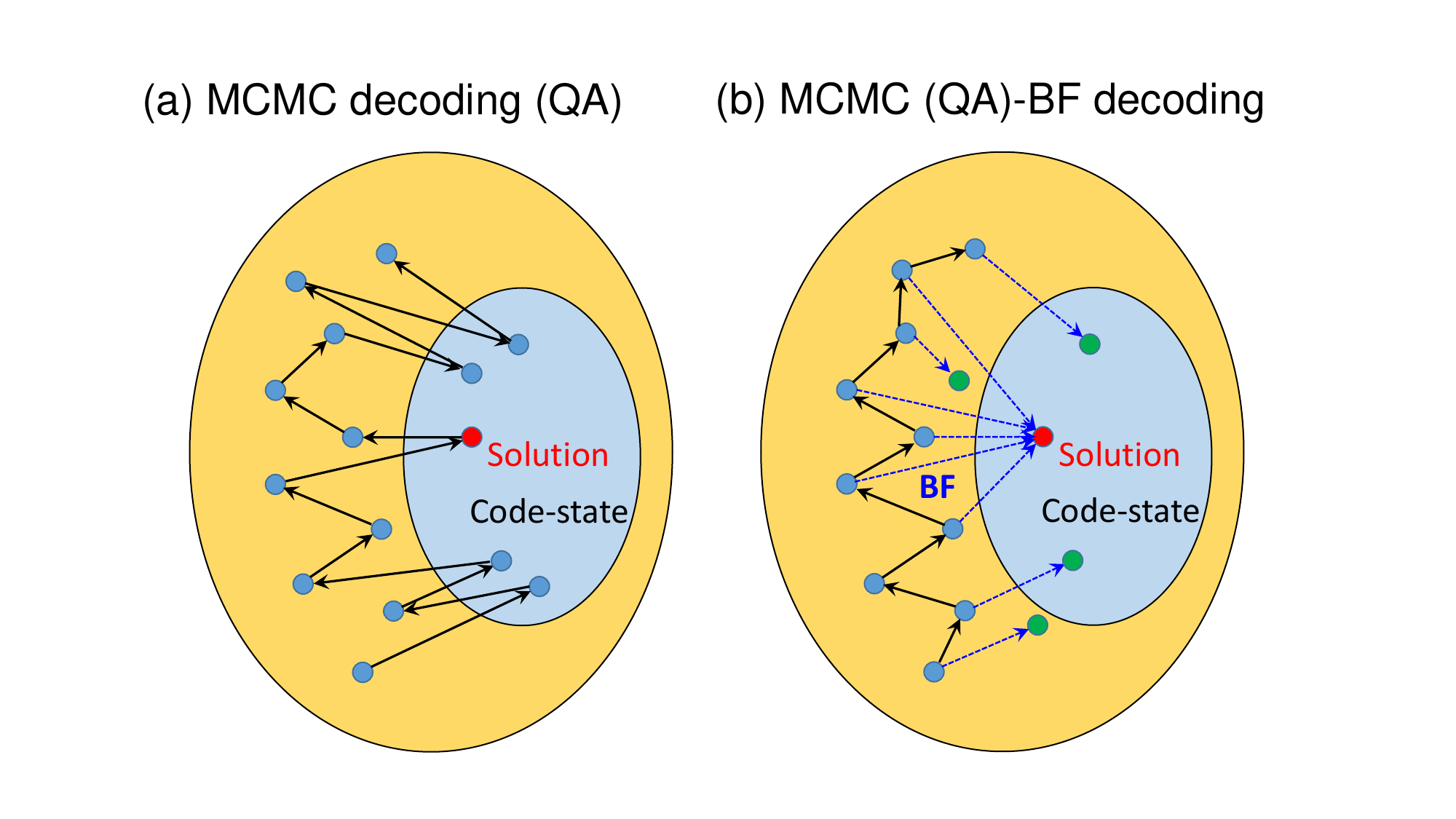}
\caption{Schematic diagrams illustrating two modes that optimize (a) MCMC decoding (and QA) and (b) MCMC(QA)-BF hybrid decoding.  The large and small ellipses indicate the search space and the code-state space, respectively. Blue circles indicate the sampled state by the MCMC or QA sampler. Green circles indicate the decoded states by the BF decoding. The red circle designates the correct code-state. Black solid arrows indicate the transitions to the following samples. Blue broken arrows indicate the BF decoding.
\label{fig:11}}
\end{figure*}

Furthermore, introducing MCMC decoding in the first stage also solves another shortcoming. The MCMC sampling is stochastic in contrast to our deterministic BF algorithm and gradient descent algorithm used in the GDBF algorithm. This stochasticity introduces randomness into the spin-flip selection. It allows spins to flip even when $\varDelta_{k}^\mathrm{(MCMC)}(\boldsymbol{x})>0$, which provides an escape from spurious local minima and makes it more likely to arrive at the neighborhood of the global minimum of $H^\mathrm{(MCMC)}(\boldsymbol{x})$. In this way, the stochastic spin-flip selection helps find the global minimum of $H^\mathrm{(MCMC)}(\boldsymbol{x})$. A similar stochasticity can be incorporated directly into the GDBF algorithm, either by adding a noise term to the inversion function $\varDelta_{k}^\mathrm{(GDBF)}(\boldsymbol{x})$ (Noisy GDBF \cite{sundararajanNoisyGradientDescent2014a}) or by taking account of stochastic spin-flip selection in the algorithm (Probabilistic GDBF \cite{rasheedFaultTolerantProbabilisticGradientDescent2014}). In other words, our MCMC-BF hybrid decoding algorithm can be considered an alternative to the BP algorithm and other existing BF decoding algorithms. In addition, the two annealing parameters $\left\{ \beta,\gamma\right\} $ in the inversion function $\varDelta_{k}^\mathrm{(MCMC)}(\boldsymbol{x})$ control the intensity of fluctuations in MCMC sampling and the relative contribution of the correlation and penalty terms. Our simulation suggests that proper control of these annealing parameters is important to optimize the performance of MCMC-BF decoding. It also suggests that the contribution of the penalty term to the Hamiltonian must be small enough not to force $\boldsymbol{x}$ into a valid code-state. Controlling annealing parameters makes it possible to switch between the two decoding modes of MCMC sampling. Fig.\ref{fig:11} shows schematic diagrams illustrating two modes that optimize (a) MCMC decoding and (b) MCMC-BF hybrid decoding, respectively. Decoding is formulated as a constrained COP. The large and small ellipses indicate the search space and the state space that minimizes the penalty term of the Hamiltonian, i.e. the code-state space. In MCMC decoding, the MCMC samples both the code-states (feasible solution) and leakage states (infeasible solution) (Fig.\ref{fig:11}(a)). In some occasions, the correct code-state (solution) may be sampled. In MCMC-BF hybrid decoding, on the other hand, the first-stage MCMC sampler samples only the leakage state, and the second-stage BF decoding maps the leakage state to the correct code-state occasionally 
(Fig.\ref{fig:11}(b)). Our simulations showed that the MCMC-BF hybrid decoding (Fig.\ref{fig:11}(b)) was about 300 times more efficient than the MCMC decoding (Fig.\ref{fig:11}(a)) if we ignore the decoding cost of the BF decoding and compare performance simply by the number of required MCMC samples needed to obtain at least one correct code-state. We will discuss the decoding cost of the BF decoding next.

\subsection{Spin-flipping mechanism and decoding cost}

The above discussions did not discuss the mechanism for choosing the spins to be flipped at each iteration, nor did they discuss the decoding cost of the BF algorithm. We want to discuss these points briefly. Spin-flipping mechanisms are free to choose from several strategies. For example, we can flip only the most suitable spin chosen based on the inversion functions $\varDelta_{k}^{(X)}(\boldsymbol{x})$'s estimated by the current decision $\boldsymbol{x}$. Alternatively, we can flip several spins chosen based on $\varDelta_{k}^{(X)}(\boldsymbol{x})$'s together. These are called the single-spin flipping and the multi-spin flipping strategies, respectively. In this classification, MCMC decoding algorithm belongs to the single-spin flipping strategy, while our BF decoding algorithm belongs to the multi-spin flipping strategy. Note that the BF decoding algorithm consists of two operations: evaluating the inversion function $\varDelta_{k}^\mathrm{(BF)}(\boldsymbol{x})$  and determining spins to be flipped. The inversion function $\varDelta_{k}^\mathrm{(BF)}(\boldsymbol{x})$  for each spin $k$  is computed at once from the current $\boldsymbol{x}$ by matrix multiplication. Subsequent sign evaluation for each element of $\varDelta_{k}^\mathrm{(BF)}(\boldsymbol{x})$ determines which spins are flipped together.

We can see from Fig.\ref{fig:2} that more edges are connected to VNs for the weight-3 syndromes ($d_{v}=K-2$) than the weight-4 syndromes ($d_{v}\leq4$). Thus, each spin variable affects more syndromes when using the weight-3 syndrome than the weight-4 syndrome. Conversely, each spin variable is determined from more syndromes when using the weight-3 syndrome than the weight-4 syndrome. Consequently, when using the weight-3 syndrome, one must compute the sum of products of data distributed globally across many spins to decide whether to invert each spin. In contrast, when using the weight-4 syndrome, it is sufficient to compute the sum of products of data distributed over at most eight adjacent spins to determine whether to invert each spin. Such a global reduction operation to compute the inversion function, which occurs when using the weight-3 syndrome in the BF decoding algorithm, requires more effort, but can accelerate the correction of errors in each spin-flip operation.

On the other hand, we can perform the matrix multiplication required for calculating $\varDelta_{k}^\mathrm{(BF)}(\boldsymbol{x})$ easily and much more benefit from parallel computing techniques and hardware engines that calculate matrix sum products, such as GPUs and vector processing engines developed for machine learning. This fact provides practical merit of choosing weight-3 syndrome for syndrome $s_{i}(\boldsymbol{x})$ in $\varDelta_{k}^\mathrm{(BF)}(\boldsymbol{x})$ (Eq.(\ref{eq:36})). In fact, comparing the computation times of the MCMC and MCMC-BF hybrid decoding presented in Fig.\ref{fig:9} performed on the Mathematica platform, the computation time of MCMC-BF hybrid decoding was one-fifth of that of the MCMC decoding. Although the computational cost depends on the algorithm used and the software or hardware platforms on which we perform the calculations, this result suggests that the MCMC-BF hybrid decoding is more efficient than MCMC decoding alone. Furthermore, we confirmed that our BF algorithm was much less computationally expensive than the BP algorithm. For example,  the computation time of the performance evaluation for the BF decoding shown in Fig.\ref{fig:6} (a) was about 1/40 of that for the BP decoding shown in Fig.\ref{fig:6}  (b). 

This paper does not discuss decoding cost in any further detail. Let us recall that the purpose of this paper is to propose and demonstrate the potential of the BF decoding as a practical post-processing algorithm for a pre-processing stochastic algorithm, such as a QA device. Since the development of QA devices is now ongoing, it is not easy to demonstrate our BF decoding algorithm as a post-processing algorithm for an actual QA device. Instead, the potential of the BF decoding was demonstrated using a classical MCMC sampler as a pre-processor. We believe that the present result is principally dependent on the property of the BF decoding algorithm, not the MCMC sampler. Furthermore, our result is consistent with the general belief that two decoding algorithms based on different mechanisms can be used together to solve problems that cannot be solved by either one alone. We believe that our BF decoding algorithm is promising for correcting errors in the readouts of QA devices due to measurement errors as well as dynamic errors that may be encountered during QA. 

\subsection{Relevance to the previous work by Albash et al.}

In Ref.\cite{albashSimulatedquantumannealingComparisonAlltoall2016}, the authors present minimum weight decoding (MWD), which aims to find the nearest Hamming distance, constraint-satisfying physical spin state from the hard-decided spin readout $\boldsymbol{x}\in\{\pm1\} ^{N_{v}}$ of the SLHZ. To this end, they directly estimated  the optimal error pattern $\boldsymbol{e}^{*}\in\{\pm1\} ^{N_{v}}$ from the global ground state of the following Hamiltonian: 
\begin{equation}
H^\mathrm{(MWD)}(\boldsymbol{w})
=-\sum_{i=1}^{N_{v}}w_{i}-\lambda \sum_{i=1}^{N_{c}} s_{i}^{(4)}(\boldsymbol{x}) s_{i}^{(4)}(\boldsymbol{w}),
\end{equation}
where $\boldsymbol{w}=(w_{1},\ldots,w_{N_{v}})\in\{\pm1\} ^{N_{v}}$  is an arbitrary spin state, and $\boldsymbol{s}^{(4)}(\boldsymbol{w})=(s_{1}^{(4)}(\boldsymbol{w}),\ldots,s_{N_{c}}^{(4)}(\boldsymbol{w}))\in\{\pm1\} ^{N_{c}}$  is the associated weight-4 syndrome vector. Here, $\lambda$ is assumed to be sufficiently large such that $\boldsymbol{w}$ satisfies all constraint terms. Let $C$ denote the set of $\boldsymbol{w}$ such that $\boldsymbol{s}^{(4)}(\boldsymbol{w})=\boldsymbol{s}^{(4)}(\boldsymbol{x})$. Then, the optimal estimate $\boldsymbol{e}^{*}$ is given by
\begin{equation}
\boldsymbol{e}^{*}=\underset{
\boldsymbol{w}\in C}
{\arg\min}H^\mathrm{(MWD)}(\boldsymbol{w}).
\end{equation}
Thus, the optimal estimate $\boldsymbol{e}^{*}$ is the error pattern with the same syndrome vector and the smallest number of errors (i.e., the smallest number of elements $e_{k}=-1$). Consequently,  the MWD decoded state $\boldsymbol{z}^{*}$ is given by $\boldsymbol{z}^{*}=\boldsymbol{x}\circ\boldsymbol{e^{*}}$. The authors also related the MWD to the ground state of the following Hamiltonian: 
\begin{equation}
H_\mathrm{spin}^\mathrm{(MWD)}(\boldsymbol{s})
=-\sum_{i<j}g_{ij}s_{i}s_{j},\label{eq:48}
\end{equation}
where a set of spin variables $\boldsymbol{g}=(g_{12},\ldots,g_{K-1\,K})\in\{\pm 1\}^{\tbinom{K}{2}}$ can be regarded as a gauge transformation. Once the ground state of Eq.(\ref{eq:48}) is found, one can deduce the optimal estimate  $\boldsymbol{e}^{*}$ and obtain the MWD decoded state $\boldsymbol{z}^{*}$. The problem here is that the MWD requires the optimal solution of the Hamiltonian defined in Eq. (\ref{eq:36}). This is equivalent to solving an instance of maximum 2-satisfiability, which is as hard as the original COP to be solved. Therefore, the MWD does not necessarily resolve the performance bottleneck problem for the SLHZ system. We think that MWD is difficult to apply in realistic scenarios when considering implementation in QA devices.

Their decoding strategy is to find an error pattern $\boldsymbol{e}$ with minimum Hamming distance from the hard-decided readouts $\boldsymbol{x}$ and with the same syndrome vector as $\boldsymbol{s}^{(4)}(\boldsymbol{x})$. In contrast, our BF decoding algorithm is an approximation, where the current hard-decided estimate  $\boldsymbol{x}$  is updated iteratively to estimate $\boldsymbol{x}'$ better. If the estimate $\boldsymbol{x}$ converges to a certain state with finite iterations, it gives a correctly decoded state  $\boldsymbol{z}^{*}$ with a finite probability. Our BF algorithm can be executed on a deterministic Turing machine in polynomial time since it is executed by finite iterations of matrix multiplication and the associated sign evaluation of the entries of the resultant matrix.  Therefore, we think our BF decoding algorithm is practical and realistic when considering implementation in QA devices.

It was pointed out that the SLHZ system faces more challenges with single spin updates than the ME system. We expect that they can be alleviated by introducing the BF decoding. Recall that, in the ME and SLHZ systems, there is an overhead resulting from the embedding, where the chain of short-range physical interactions simulates the effect of long-range physical interactions. Although a chain consists of local two-body interactions in the ME system, it consists of local four-body interactions in the SLHZ system. On the other hand, the spin update in our BF decoding algorithm relies on numerous weight-3 syndromes corresponding to long-range three-body interactions. As shown in Fig.\ref{fig:2}, the column weight $d_{v}$ for the weight-4 syndrome is always less than 4, while that for the weight-3 syndrome depends on the size $K$ of the original logical problem and increases with increasing $K$. 
For $K>6$, the use of the weight-3 syndrome instead of the weight-4 syndrome can make spin updates more efficient since it increases the connectivity between spins. This was actually confirmed in the following observation.  Let us recall that MCMC decoding using the Hamiltonian (\ref{eq:32}) results in the performance curves shown in Fig \ref{fig:6}(c). However, when the same evaluation was performed by replacing the weight-3 syndrome $s_{ijk}^{(3)}(\hat{\boldsymbol{x}})$ with the weight-4 syndrome $s_{ijk}^{(4)}(\hat{\boldsymbol{x}})$, the performance was found to drop significantly. This is quite reasonable, considering that the mixing property of the MCMC depends on connectivity between the spins; the more connectivity each spin has, the better the mixing property should be. In other words, embedding is not possible without sacrificing mixing properties. Since we don't have to worry about technical restrictions on connectivity if we use a digital computer, we can use weight-3 syndromes for the post-readout BF decoding. Therefore, although the spin-update of SLHZ system is less efficient, the post-readout BF decoding can vastly improve its spin-update properties. 

Our BF algorithm lacks consideration of soft information related to the logical coupling constants $\boldsymbol{J}=(J_{12},\ldots,J_{K-1\,K})\in\mathbb{R}^{\tbinom{K}{2}}$. Thus, it is important to increase the relative weight of the correlation term to the penalty term to recover the soft information contained in the correlation term of the Hamiltonian $H^{code}\left(\hat{\boldsymbol{x}}\right)$ of the SLHZ system given by Eq.(34). Previously, Albash et al. had tested with relevant distributions generated using simulated quantum annealing (SQA) and parallel tempering (PT) and found that BP, as well as MWD, offer a substantial performance boost over majority vote decoding (MVD) when the penalty strength is brought close to zero. However, they thought that this boost was totally due to MWD itself, because it was typically seen in a specific class of instances where the ground state of the original logical problem that was approximated by $J_{ij}\to\mathrm{sign}(J_{ij})$ and the ground state of the original logical problem happen to coincide. On the other hand, the similar performance boost was also observed for the BF decoding when the penalty strength was almost zero as shown in Fig.\ref{fig:9}. In our simulations, similar characteristics were observed in all instances studied, regardless of whether they belonged to the above specific class. We believe this performance boost comes from the mechanism schematically depicted by Fig.\ref{fig:11}  in  \ref{subsec:5-A}. Since our BF algorithm ignores the soft information related to the logic coupling constant J and uses only the weight-3 syndrome as a criterion for updating the spin state, to be decoded to the correct state, the spin state just before it is applied must be in a state with correctable leakage errors. By controlling the penalty strength, the soft information contained in the correlation term of the Hamiltonian $H^{code}\left(\hat{\boldsymbol{x}}\right)$ of the  SLHZ system can be adequately reflected in the readout state, boosting the sampling probability of the states with correctable leakage errors.

As seen from the above, our study suggests that even though the BF and BP algorithms were initially designed to deal with independent or weakly correlated spin-flip errors, they can also be used to deal with correlated errors in SLHZ systems. It also suggests that appropriately controlling the strength of the penalty is crucial to exploiting the full potential of the fault tolerance inherent in the SLHZ system. We recognize the performance boost at vanishing penalty strength as a feature specific to the SLHZ system that shows intrinsic fault tolerance against correlated errors. Here, let us recall that the main focus of this study is to propose a practical decoding algorithm that can handle correlated spin-flip errors. We do not claim that this performance boost is sufficient for the SLHZ system to outperform the ME system, but we do consider that it is at least essential for the SLHZ system to be competitive with the ME system. We confirmed that this is indeed required in our preliminary experiments using the MCMC sampler and that ME system shows no performance boost at vanishing penalty strength even after post-readout MVD. 

\section{Conclusions\label{sec:6}}

This paper studied a practical decoding algorithm for correcting errors in the readout of the spins in the SLHZ systems. Given the close relationship between COP based on the SLHZ system and the decoding of the classical LDPC codes associated with the AWGN channel model, classical decoding techniques for LDPC codes are expected to provide noise immunity for the SLHZ system. 
We proposed a BF decoding algorithm based on iterative majority voting, which is known as a decoding method for LDPC codes and is much simpler than the standard BP algorithm. We demonstrated that the BF decoding algorithm provides good protection against i.i.d. noise in the spin readout of the SLHZ system as efficiently as the BP decoding algorithm. To study the tolerance to errors caused by noise other than i.i.d. noise, we conducted a classical simulation of spin readouts on the SLHZ system using MCMC sampler. We found that BF decoding can correct leakage errors due to classical thermal noise. An efficient decoding is possible if we reduce energy penalties so that thermal excitations are more likely to populate correctable leakage states. This implies that the SLHZ system exhibits fault tolerance against a broader range of noise models than i.i.d. noise model if collaborating with the post-readout BF decoding. Our observation is quite reasonable if we note that stochastic sampling followed by error correction can be regarded as a two-stage hybrid decoding algorithm. Controlling the annealing parameters in the first-stage sampler is essential to obtain the correct state efficiently. Special attention should be paid to the fact that the optimal annealing parameters depend heavily on whether or not decoding is performed on the sampled readout.

Since our demonstration used readouts stochastically sampled by a classical method, specifically MCMC sampling, we are not fully convinced that the BF decoding is also valid for readouts obtained through QA device using the SLHZ system. Further research is needed to assess how effectively the QA device performs collaboratively with BF decoding under realistic laboratory conditions. Nevertheless, we believe that most of our insights stem from the intrinsic nature of BF decoding and are applicable regardless of the stochastic sampling mechanism employed in the initial stage. For example, if measurement error is the primary source of error in the spin readout of the SLHZ system, our BF decoding algorithm offers a straightforward solution to mitigate it. In addition, it would be reasonable to expect that a two-stage hybrid computation combining QA and post-readout BF decoding may address issues neither method can solve independently. Our research also emphasizes the importance of adequately selecting decoding algorithms to exploit the potential of fault tolerance inherent in SLHZ systems. 

In this study, we tested only a small number of  instances $K_{14}$ with fixed-size to investigate the performance and characteristics of the BF decoding, which is insufficient to draw a general conclusion. Of particular importance for future work is investigating the size dependence on the performance of the SLHZ system. In the present discussion, we suggested that the post-readout BF decoding may compensate for the poor spin-update properties of the SLHZ system to some extent. It will be interesting to see how this compensation works well for a problem of arbitrary size. Albash et al. also pointed out another important limitation of the SLHZ system: the strength of the energy penalty must grow with the problem size \cite{albashSimulatedquantumannealingComparisonAlltoall2016}. Introducing the post-readout BF decoding may alleviate this limitation since it reduces the penalty strength. Both of these are issues that need to be studied to realize scalability. More research is needed to determine the efficacy of post-readout decoding. While the performance evaluation using the states generated by PT and SQA is reasonable, questions remain about the performance evaluation based on the number of sweeps necessary for generating them. This is because MC describes spin dynamics as a sequential update of spin states at each sweep, which is a serial and discrete time evolution. In contrast, QA dynamics always involve a parallel and continuous time evolution of spins. Therefore, the number of MC sweeps required and QA performance are not necessarily correlated. A reasonable performance evaluation would need to await the development of an actual QA device.

\begin{acknowledgments}
I would like to thank Dr. T. Kadowaki at Global Research and Development Center for Business by Quantum-AI technology (G-QuAT) and Prof. H. Nishimori at Institute of Science Tokyo for their useful comments and discussions. I also thank Dr. Masayuki Shirane of NEC Corporation/National Institute of Advanced Industrial Science and Technology for his continuous support. This paper is partly based on results obtained from a project, JPNP16007, commissioned by the New Energy and Industrial Technology Development Organization (NEDO), Japan. 
\end{acknowledgments}

\nocite{*}
\bibliography{APS}

\end{document}